\documentclass[plaindraft]{jpp-AAS}

\usepackage{graphicx,amsmath}
\usepackage{natbib}
\usepackage{hyperref}
\hypersetup{colorlinks,allcolors=blue}
\usepackage{cleveref}

\def\apjs{Astrophys.~J.~Supp.~Ser.}
\def\apj{Astrophys.~J.}
\def\apjl{Astrophys.~J.~Lett.}

\def\mnras{Mon.~Not.~Roy.~Astron.~Soc.}

\def\prl{Phys.~Rev.~Lett.}
\def\jgr{J.~Geophys.~Res.}

\def\grl{Geophys.~Res.~Lett.}
\def\pop{Phys.~Plasmas}
\def\pof{Phys.~Fluids}
\def\pofb{Phys.~Fluids~B}
\def\pnas{Proc.~Nat.~Acad.~Sci.}
\def\jcp{J.~Comput.~Phys.}
\def\jpp{J.~Plasma Phys.}

\def\ropp{Rev.~Plasma Phys.}

\def\njp{New J.~Phys.}

\ifCUPmtlplainloaded \else
  \checkfont{eurm10}
  \iffontfound
    \IfFileExists{upmath.sty}
      {\typeout{^^JFound AMS Euler Roman fonts on the system,
                   using the 'upmath' package.^^J}%
       \usepackage{upmath}}
      {\typeout{^^JFound AMS Euler Roman fonts on the system, but you
                   dont seem to have the}%
       \typeout{'upmath' package installed. JPP.cls can take advantage
                 of these fonts, if you use 'upmath' package.^^J}%
       \providecommand\upi{\pi}%
      }
  \else
    \providecommand\upi{\pi}%
  \fi
\fi

\ifCUPmtlplainloaded \else
  \checkfont{msam10}
  \iffontfound
    \IfFileExists{amssymb.sty}
      {\typeout{^^JFound AMS Symbol fonts on the system, using the
                'amssymb' package.^^J}%
       \usepackage{amssymb}%
       \let\le=\leqslant  
       \let\ge=\geqslant  
      }{}
  \fi
\fi

\ifCUPmtlplainloaded \else
  \IfFileExists{amsbsy.sty}
    {\typeout{^^JFound the 'amsbsy' package on the system, using it.^^J}%
     \usepackage{amsbsy}}
    {\providecommand\boldsymbol[1]{\mbox{\boldmath $##1$}}}
\fi

\newcommand{\pD}[2]{\frac{\partial #2}{\partial #1}}
\newcommand{\pDD}[2]{\frac{\partial^2 #2}{\partial {#1}^2}}
\newcommand{\D}[2]{\frac{{\rm d} #2}{{\rm d} #1}}
\newcommand\bb[1]{\mbox{\boldmath{$#1$}}}
\newcommand{\circleit}[1]{\raisebox{0.05em}{$\bigcirc$}$\hspace{-0.76em} {#1}$\;}

\newcommand{\imag}{{\rm i}}
\newcommand{\rmd}{{\rm d}}
\newcommand{\rme}{{\rm e}}
\newcommand{\ex}{\hat{\bb{x}}}
\newcommand{\fMi}{f_{{\rm M},i}}
\newcommand{\vth}[1]{v_{{\rm th}#1}}
\newcommand{\betaio}{\beta_{i0}}

\title[Self-sustaining sound in high-$\beta$ plasma]{Self-sustaining sound in collisionless, \\ high-$\bb{\beta}$ plasma}

\author[M.~W.~Kunz and others]%
{M.~W.~Kunz\ls$^{1,2}$%
  \thanks{Email address for correspondence: mkunz@princeton.edu}, 
J.~Squire$^{3}$, 
A.~A.~Schekochihin$^{4,5}$, and \\
E.~Quataert$^{6}$}

\affiliation{$^1$Department of Astrophysical Sciences, Princeton University, Peyton Hall, Princeton, NJ 08544, USA\\[\affilskip]
$^2$Princeton Plasma Physics Laboratory, PO Box 451, Princeton, NJ 08543, USA\\[\affilskip]
$^3$Department of Physics, University of Otago, 730 Cumberland St, North Dunedin, Dunedin 9016, New Zealand\\[\affilskip]
$^4$Rudolf Peierls Centre for Theoretical Physics, University of Oxford, Clarendon Laboratory, Parks Road, Oxford OX1 3PU, UK\\[\affilskip]
$^5$Merton College, Merton Street, Oxford OX1 4JD, UK\\[\affilskip]
$^6$Department of Astronomy and Theoretical Astrophysics Center, University of California, Berkeley, Berkeley, CA 94720, USA}

\pubyear{2020}
\volume{}
\pagerange{}
\date{\today}

\begin{document}

\maketitle

\begin{abstract}
Using analytical theory and hybrid-kinetic numerical simulations, we demonstrate that, in a collisionless plasma, long-wavelength ion-acoustic waves (IAWs) with  amplitudes $\delta n/n_0 \gtrsim 2/\beta$ (where $\beta\gg{1}$ is the ratio of thermal to magnetic pressure) generate sufficient pressure anisotropy to destabilize the plasma to firehose and mirror instabilities. These kinetic instabilities grow rapidly to reduce the pressure anisotropy by pitch-angle scattering and trapping particles, respectively, thereby impeding the maintenance of Landau resonances that enable such waves' otherwise potent collisionless damping. The result is wave dynamics that evince a weakly collisional plasma: the ion distribution function is near-Maxwellian, the field-parallel flow of heat resembles its Braginskii form (except in regions where large-amplitude magnetic mirrors strongly suppress particle transport), and the relations between various thermodynamic quantities are more `fluid-like' than kinetic. A nonlinear fluctuation-dissipation relation for self-sustaining IAWs is obtained by solving a plasma-kinetic Langevin problem, which demonstrates suppressed damping, enhanced fluctuation levels, and weakly collisional thermodynamics when IAWs with $\delta n/n_0 \gtrsim 2/\beta$ are stochastically driven. We investigate how our results depend upon the scale separation between the wavelength of the IAW and the Larmor radius of the ions, and discuss briefly their implications for our understanding of turbulence and transport in the solar wind and the intracluster medium of galaxy clusters.
\end{abstract}

%
%
\section{Introduction}

The material properties of high-$\beta$ plasmas are complicated greatly by the magnetic fields they harbor.\footnote{$\beta \doteq 8\upi p/B^2$ is the ratio of the thermal pressure of the plasma particles, $p$, and the energy density stored in the magnetic field, $B^2/8\upi$. In this paper, `high-$\beta$' means $\beta \gg 1$.} This is particularly true in hot, dilute astrophysical plasmas such as the intracluster medium (ICM) of galaxy clusters, some regions of the solar wind, and certain classes of accretion flows onto supermassive black holes. These plasmas exhibit enormous scale separation between the gyroradii of their constituent particles and macroscopic scales, such as the system size, temperature- and momentum-gradient lengthscales, and collisional mean free paths. The result is that magnetic fields, even when energetically subdominant, can nevertheless change the transport properties of a plasma in a way that profoundly influences the large-scale dynamics it hosts. Examples of this include the fact that anisotropic (i.e., field-aligned) conduction and viscosity fundamentally change the properties of convective and rotational instabilities that are critical for heat and momentum transport in a wide variety of astrophysical plasmas \citep[e.g.][]{balbus00,balbus01,qdh02,balbus04,quataert08,kunz11,xk16}.

In this paper, we highlight another way in which energetically weak magnetic fields change the transport properties of a low-collisionality, magnetized plasma. As an ion-acoustic wave (IAW) propagates through a magnetized plasma, velocity-space anisotropy in the particle distribution function is driven by anisotropic phase mixing and (approximate) adiabatic invariance. For relative wave amplitudes larger than ${\sim}2/\beta$, this anisotropy triggers rapidly growing, Larmor-scale kinetic instabilities (firehose, mirror). These instabilities ultimately scatter and/or trap particles, thereby interfering with Landau resonances and thus stifling the otherwise potent collisionless damping that linear IAWs would suffer. As a result, IAWs of sufficiently large amplitude can be self-sustaining, and propagate in a manner akin to sound waves in a weakly collisional fluid. This result complements recent work showing that low-collisionality plasmas cannot support linearly polarized shear-Alfv\'{e}n fluctuations above a critical $\beta$-dependent amplitude \citep{squire16,squire17num,squire17}.

%
%
\section{Prerequisites: pressure anisotropy, wave  `interruption', and collisionless damping}

For our purposes, the qualifier `low-collisionality' means that particle--particle collisions are sufficiently rare for each particle's adiabatic invariants to be approximately conserved. In a low-collisionality plasma, changes in magnetic-field strength $B$ and density $n$ are thus accompanied by changes in the perpendicular ($\perp$) and parallel ($\parallel$) pressures of the plasma: $p_\perp \propto nB$ and $p_\parallel \propto n^3/B^2$  \citep{cgl56}. In other words, the energies in the random motions of the constituent particles in the directions perpendicular and parallel to the local magnetic field become different -- the pressure becomes {\em anisotropic}, with $p_\perp \ne p_\parallel$. In a high-$\beta$ plasma, even small pressure anisotropy $\Delta \doteq p_\perp/p_\parallel - 1$ can drive rapidly growing ion-Larmor-scale instabilities, most notably the firehose (when $\Delta \lesssim -2/\beta$; \citealt{rosenbluth56,ckw58,parker58,vs58,yoon93,hm00}) and the mirror (when $\Delta \gtrsim 1/\beta$; \citealt{barnes66,hasegawa69,sk93,hellinger07}). Recent numerical work \citep{kss14,riquelme15,msk16,ht15,sn15,hellinger17,riquelme18} has shown that these instabilities effectively regulate the pressure anisotropy to marginally stable values, in agreement with solar-wind observations \citep{kasper02,hellinger06,bale09,chen16}. 

Marginal firehose stability is achieved by generating electromagnetic fluctuations on ion-Larmor scales, which ultimately pitch-angle scatter particles at the required rate. This severs the adiabatic link between large-scale decreases in magnetic-field strength or increases in plasma density and the production of negative pressure anisotropy, $\Delta < 0$. In other words, the firehose instability provides the otherwise collisionless plasma with just enough `collisionality' to limit its deviations from local thermodynamic equilibrium. In the case of the mirror instability, this adiabatic link is exploited rather than severed: as the pressure anisotropy $\Delta$ increases and the system goes mirror unstable, an increasing population of resonant particles become trapped in deepening magnetic troughs, thereby cooling perpendicularly to reduce the total $\Delta$. In effect, the mirror instability `hides' the large-scale adiabatic production of pressure anisotropy from a majority of the plasma particles by corralling them into small-scale safe spaces, absolving them of any need to adjust their velocity distribution \citep{schekochihin08,kss14,rincon15,msk16}.

The pressure anisotropies inevitably driven (linearly or non-linearly) by large-scale waves in low-collisionality plasmas, and the kinetic instabilities that reign them in, change the wave dynamics of high-$\beta$ plasmas in remarkable ways. In a recent series of publications, \citet{squire16,squire17num,squire17} have shown that, in collisionless plasmas, linearly polarized shear-Alfv\'{e}n waves with amplitudes $\delta B_\perp / B_0 \gtrsim \beta^{-1/2}$ are strongly modified by the pressure anisotropies that they produce nonlinearly. Larger wave amplitudes lead to larger wave-generated pressure anisotropy, eventually reaching the point at which the pressure anisotropy exactly offsets the restoring Lorentz force (which is the same as reaching the firehose instability threshold). This nullifies the restoring force for the Alfv\'{e}n wave and inhibits its propagation, an effect those authors termed {\em wave interruption}.  Accompanying this interruption is the emergence of a sea of ion-Larmor-scale mirror and firehose fluctuations, which respectively trap and pitch-angle scatter particles and thus provide an effective plasma viscosity.

What of {\em compressive} fluctuations? Consider a long-wavelength, parallel-propagating IAW in a collisionless, magnetized, ion--electron plasma. Assuming the equilibrium distribution functions of the ions and electrons to be Maxwellians with temperatures $T_i$ and $T_e$, respectively, and neglecting terms that are small in square root of the mass ratio $(m_e/m_i)^{1/2}$, one finds the dispersion relation
\begin{equation}\label{eqn:IAW}
    D(\zeta) \doteq 1 + \frac{T_i}{T_e} + \zeta Z(\zeta) = 0,
\end{equation}
where $\zeta \doteq (\omega-\imag\gamma)/|k_\parallel|\vth{i}$ is the dimensionless phase speed of a fluctuation with (real) frequency $\omega$, decay rate $\gamma$ $(>0)$, and parallel wavenumber $k_\parallel$; $Z(\zeta)$ is the plasma dispersion function; and $\vth{i} \doteq (2T_i/m_i)^{1/2}$ is the ion thermal speed. `Long-wavelength' means $|k_\parallel| \rho_i \ll 1$, where $\rho_i \doteq \vth{i}/\Omega_i$ is the ion Larmor radius and $\Omega_i$ is the ion Larmor frequency. In the limit $T_i/T_e\ll{1}$, equation \eqref{eqn:IAW} has the well-known solution
\begin{equation}\label{eqn:IAW_wk}
    \frac{\omega}{|k_\parallel|\vth{i}} \approx \biggl(\frac{T_e}{2T_i}\biggr)^{1/2} , \quad \frac{\gamma}{|k_\parallel|\vth{i}} \approx \sqrt{\upi} \, \biggl(\frac{T_e}{2T_i}\biggr)^2 \exp\biggl(-\frac{T_e}{2T_i}\biggr) ,
\end{equation}
describing weakly damped IAWs propagating in the direction of the magnetic field \citep{bk60,fg61}. When the ion and electron temperatures are comparable, the decay rate is instead comparable to the oscillation frequency, i.e., $\gamma \sim \omega \sim |k_\parallel|\vth{i}$.
%
%

For appreciable collisionless damping to occur, it is necessary that the particles be able to stream freely, so that they can both maintain Landau resonance with the wave and phase mix the perturbed distribution function. In a collisional (i.e., magnetohydrodynamic (MHD)) plasma, these processes do not occur, because particle trajectories are randomized by frequent particle--particle collisions. With $|k_\parallel|\lambda_{\rm mfp,c} \ll 1$ in the MHD limit, where $\lambda_{\rm mfp,c}$ is the Coulomb-collisional mean-free path, small-amplitude sound waves propagate just fine.

In the next section, we use analytical theory and numerical simulations to investigate how velocity-space anisotropy generated by long-wavelength, finite-amplitude IAWs influences their propagation and damping. We show that, above a $\beta$-dependent amplitude threshold, the free streaming of particles required for collisionless damping to occur is impeded by firehose and mirror instabilities that are driven by the wave-generated pressure anisotropy. The behaviour of the wave becomes MHD-like.

%
%
\section{Interruption of the Landau damping of IAWs}

\subsection{Theory}\label{sec:IAWtheory}

Suppose that the ion distribution function $f_i(t,\bb{r},\bb{v})$ of a collisionless, magnetized plasma threaded by a mean magnetic field $\bb{B}_0 = B_0 \ex$ is given at $t=0$ by
\begin{equation}\label{eqn:f0}
    f_i(0,x,v) = \frac{n_0 + \delta n(0,x)}{\upi^{3/2} \vth{i0}^3} \exp\left( - \frac{v^2}{\vth{i0}^2} \right) =  \left[ 1 + \frac{\delta n(0,x)}{n_0} \right] \fMi(v),
\end{equation}
where $\delta n(0,x)=\alpha\,n_{0}\cos(K x)$ is a small ($\alpha\ll{1}$), long-wavelength ($K\rho_i\ll{1}$) perturbation to the mean number density $n_0$, $\vth{i0}=(2T_{i0}/m_i)^{1/2}$ with $T_{i0}$ being the initial ion temperature, and $\fMi(v)$ is a standard Maxwell--Boltzmann distribution with $\int\rmd^3\bb{v}\,\fMi(v) = n_0$ and $\int\rmd^3\bb{v}\, m_i v^2 \fMi(v) = 3 n_0 T_{i0} \doteq 3p_{i0}$. The perturbed piece of (\ref{eqn:f0}), {\em viz.}, $\delta f_i(0,x,v) = [\delta n(0,x)/n_0] \fMi(v)$, may be written in Fourier space as
\begin{equation}\label{eqn:dfk}
    \delta f_i(0,k_\parallel,v) = \frac{\alpha}{2}\bigl[ \delta(k_\parallel-K) + \delta(k_\parallel+K) \bigr] \fMi(v) \doteq \frac{\delta n(0,k_\parallel)}{n_0} \, \fMi(v) ,
\end{equation}
where the subscript `$\parallel$' denotes the component oriented along the direction of the mean magnetic field. Assuming for simplicity that the electrons constitute a massless, neutralizing, isothermal fluid with temperature $T_e$, such a perturbation will produce an electrostatic response,
\begin{equation}\label{eqn:varphi}
    \frac{e\varphi(t,k_\parallel)}{T_e} = \frac{\delta n(t,k_\parallel)}{n_0} = \frac{1}{n_0} \int\rmd^3\bb{v} \, \delta f_i(t,k_\parallel,v) ,
\end{equation}
which will drive parallel flows in the plasma and cause an IAW to oscillate and be damped collisionlessly at the rate determined by (\ref{eqn:IAW}). As the perturbed density initially relaxes, pressure anisotropy with respect to the magnetic-field direction ($\delta{p_\perp}\ne\delta{p_\parallel}$) will be driven by anisotropic phase mixing and adiabatic invariance.

\subsubsection{Generation of pressure anisotropy and triggering of kinetic instabilities}

We demonstrate this quantitatively by considering the linearized Vlasov equation for the ions, 
\begin{equation}\label{eqn:linvlasov}
\biggl( \pD{t}{} + \imag k_\parallel v_\parallel \biggr) \delta f_i(t,k_\parallel,v) + \frac{e\varphi(t,k_\parallel)}{T_{i0}} \imag k_\parallel v_\parallel \fMi(v) = 0 ,
\end{equation}
with the potential \eqref{eqn:varphi} and integrating it forward in time from the initial condition \eqref{eqn:dfk}. The result is
\begin{equation}\label{eqn:df0}
    \delta f_i(t,k_\parallel,v) = \fMi(v) \,\rme^{-\imag k_\parallel v_\parallel t}  \left[ \frac{\delta n(0,k_\parallel)}{n_0} - \imag k_\parallel v_\parallel \int^t_0 \rmd t' \, \rme^{\imag k_\parallel v_\parallel t'}  \frac{T_e}{T_{i0}} \frac{\delta n(t',k_\parallel)}{n_0} \right] .
\end{equation}
The first term represents phase mixing of the initial perturbation due to particles free streaming along the (unperturbed) magnetic field. The second term includes the self-consistent response; it may be manipulated, after integration by parts, to help recast \eqref{eqn:df0} as
\begin{align}\label{eqn:df1}
    \delta f_i(t,k_\parallel,v) + \fMi(v)\frac{T_e}{T_{i0}} \frac{\delta n(t,k_\parallel)}{n_0}   &= \fMi(v) \,\rme^{-\imag k_\parallel v_\parallel t}\left(1+\frac{T_e}{T_{i0}}\right) \frac{\delta n(0,k_\parallel)}{n_0}  \nonumber\\*
    \mbox{} &+ \fMi(v) \int^t_0 \rmd t' \, \rme^{-\imag k_\parallel v_\parallel (t-t')} \frac{T_e}{T_{i0}} \pD{t'}{} \frac{\delta n(t',k_\parallel)}{n_0} .
\end{align}
Taking the difference of the $m_i v^2_\perp/2$ and $m_i v^2_\parallel$ moments of \eqref{eqn:df1} eliminates the second term on the left-hand side, leaving the following expression for the ion pressure anisotropy driven by the IAW:
\begin{align}\label{eqn:IAWdelta}
    \Delta_{\rm IAW}(t,k_\parallel) &= 2 \left(\frac{k_\parallel \vth{i0} t}{2}\right)^2 \rme^{-(k_\parallel \vth{i0} t/2)^2} \left(1+\frac{T_e}{T_{i0}}\right) \frac{\delta n(0,k_\parallel)}{n_0}  \nonumber\\*
    \mbox{} &+ 2 \int^t_0 \rmd t' \, \left[\frac{k_\parallel \vth{i0} (t-t')}{2}\right]^2 \rme^{-[k_\parallel \vth{i0} (t-t')/2]^2} \, \frac{T_e}{T_{i0}}\pD{t'}{}\frac{\delta n(t',k_\parallel)}{n_0} .
\end{align}
All terms involving the combination $k_\parallel \vth{i0}t/2$ describe the damping effect of phase mixing on the moments of the perturbed distribution function due to the production of fine-scale structure along $v_\parallel$. The first term on the right-hand side of \eqref{eqn:IAWdelta} arises because no accompanying fine-scale structure is produced along $v_\perp$ -- magnetized particles cannot stream freely across the magnetic field -- and so the $v^2_\perp$ and $v^2_\parallel$ moments of $\delta f_i$ phase mix differently. (Equivalently, the parallel flows of perpendicular and parallel heat are different.) The result is a transiently produced positive pressure anisotropy. The last term on the right-hand side of \eqref{eqn:IAWdelta} captures the pressure anisotropy driven by adiabatic invariance as the density fluctuates in time. 

\begin{figure}
    \centering
    \includegraphics[width=\textwidth]{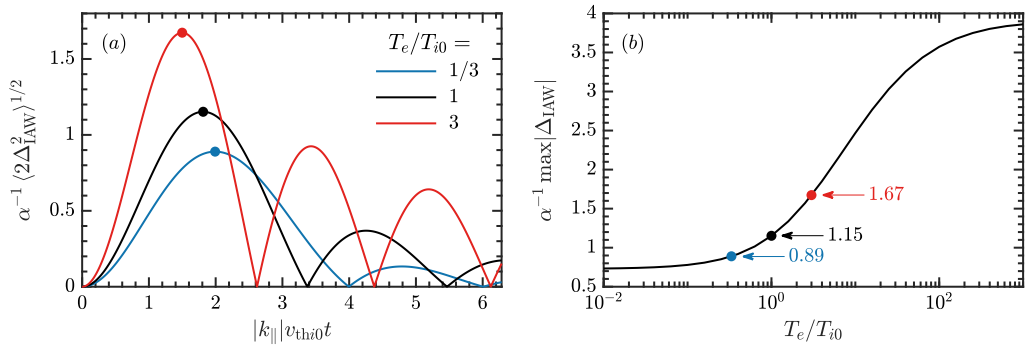}
    \caption{($a$) Solution of \eqref{eqn:IAWdelta} for the time-dependent root-mean-square pressure anisotropy of a linear IAW of wavenumber $k_\parallel$ and dimensionless amplitude $\alpha$ for various $T_e/T_{i0}$. ($b$) Maximum pressure anisotropy (divided by $\alpha$) vs.~$T_e/T_{i0}$; its values at $T_e/T_{i0}=1/3$, $1$, and $3$ are indicated.}
    \label{fig:iaw_lin}
\end{figure}

The integral in \eqref{eqn:IAWdelta} can be straightforwardly done numerically. The resulting evolution of the (spatial) root-mean-square pressure anisotropy, $\langle\Delta^2_{\rm IAW}(t,x)\rangle^{1/2}$, is shown in figure~\ref{fig:iaw_lin}($a$) at a selection of values of $T_e/T_{i0}$; the maximum value of the pressure anisotropy is plotted as a function of $T_e/T_{i0}$ in figure~\ref{fig:iaw_lin}($b$).\footnote{A multiplicative factor of $\sqrt{2}$ has been applied to $\langle\Delta^2_{\rm IAW}\rangle^{1/2}$ in figure~\ref{fig:iaw_lin}($a$) to compensate for having  $\langle\cos^2(Kx)\rangle^{1/2}=1/\sqrt{2}$. This makes it easier to read off the amplitude of $\Delta_{\rm IAW}$, which is the decisive quantity given an initial IAW amplitude $\alpha$. A similar adjustment has been applied to the root-mean-square density fluctuation in figures~\ref{fig:iaw_damping}($a$) and \ref{fig:nuscan}($c$)}.
%
%
As the IAW begins to oscillate and decay, regions with $\delta n(0,x) > 0$ ($<0$) initially drive $\Delta_{\rm IAW}(t,x) > 0$ ($<0$) towards the mirror (firehose) threshold. Once $\Delta_{\rm IAW}(t,x) \gtrsim 1/\beta_i$ ($\lesssim -2/\beta_i$), those regions become mirror (firehose) unstable. Noting that the maximum value of $\Delta_{\rm IAW}$ is a relatively weak function of $T_e/T_{i0}$, we arrive at an amplitude threshold for an IAW to trigger both instabilities:
\begin{equation}\label{eqn:IAWlimit}
    \alpha = \left|\frac{\delta n}{n_0}\right| \gtrsim \frac{2}{\beta_i} \quad \textrm{(IAW amplitude threshold)}.
\end{equation}
For example, for $T_e/T_{i0}=1$, figure~\ref{fig:iaw_lin}($b$) indicates that $\alpha = 1.74/\beta_i$ is required for ${\rm max}|\Delta_{\rm IAW}|=2/\beta_i$. Because $\beta_i \gg 1$ in many astrophysical plasmas, even small-amplitude fluctuations ({\em viz.}, those with $\alpha\ll{1}$ but $\alpha\beta_i \gtrsim 2$) generate enough pressure anisotropy to destabilize the plasma. This is the first key result of this paper.

%
%
\subsubsection{Evolution of IAW-driven kinetic instabilities and regulation of pressure anisotropy}\label{sec:IAWtheory_inst}

For initial density perturbations satisfying \eqref{eqn:IAWlimit}, the times at which the plasma becomes firehose and mirror unstable -- $t_{\rm f}$ and $t_{\rm m}$, respectively -- may be estimated by asking when the corresponding (approximate) instability parameters, $\Lambda_{\rm f} \doteq -(\Delta + 2/\beta_i)$ and $\Lambda_{\rm m} \doteq \Delta - 1/\beta_i$, pass through zero to become positive. Assuming that the combination $\alpha\beta_i$ is sufficiently large so that these times are ${\ll}(|k_\parallel| \vth{i0})^{-1}$, equation \eqref{eqn:IAWdelta} may be used to approximate the time evolution of the pressure anisotropy as $|\Delta(t)| \sim \alpha (k_\parallel \vth{i0} t)^2$. Then, neglecting factors of order unity,
\begin{equation}\label{eqn:tFHMR}
    t_{\rm f}, t_{\rm m} \sim \left( |k_\parallel| \vth{i0} \sqrt{\alpha\beta_i} \right)^{-1} .
\end{equation}
For $\alpha\beta_i \gtrsim 2$, this is a fraction of the sound-crossing time ${\sim}2\upi/|k_\parallel|\vth{i0}$.

After the plasma is driven firehose/mirror unstable (i.e., for $t>t_{\rm f},t_{\rm m}$), the pressure anisotropy will continue to grow in amplitude following \eqref{eqn:IAWdelta} until the instabilities are able to deplete the pressure anisotropy faster than it is being supplied. To determine the times at which this occurs, we first note that the maximum growth rates of the oblique firehose and mirror instabilities at high $\beta_i$, {\it viz.}~$\gamma_{\rm f} \sim \Omega_i \Lambda^{1/2}_{\rm f}$ and $\gamma_{\rm m} \sim \Omega_i \Lambda^2_{\rm m}$, respectively \citep[e.g.,][]{yoon93,hellinger07}, depend on time through $\Delta = \Delta_{\rm IAW}(t)$.\footnote{For small $\Lambda_{\rm f}$, the maximum growth rate of the parallel firehose is a factor ${\sim}\Lambda^{1/2}_{\rm f}$ smaller than its oblique counterpart \citep[e.g.,][]{ks67,dv68}.} For times such that $(t/t_{\rm f}-1), (t/t_{\rm m}-1) \ll{1}$ and that the instability parameters are both ${\ll}1$, we can obtain simple analytical expressions for these time-dependent growth rates by Taylor expanding the instability parameters about $t_{\rm f}$ and $t_{\rm m}$ and using $|\Delta(t)|\sim\alpha(k_\parallel\vth{i0}t)^2$. This gives
\begin{subequations}
\begin{align}
    \gamma_{\rm f}(t) &\sim \Omega_i \biggl(\frac{\alpha}{\beta_i}\biggr)^{1/4} (|k_\parallel| \vth{i0})^{1/2} (t-t_{\rm f})^{1/2} , \\
    \gamma_{\rm m}(t) &\sim \Omega_i \biggl(\frac{\alpha}{\beta_i} \biggr) (k_\parallel \vth{i0})^2 (t-t_{\rm m})^2 .
\end{align}
\end{subequations}
(Recall that $k_\parallel$ refers to the parallel wavenumber of the IAW, not of the kinetic instabilities.) The time-dependent amplification factors of the energies of the microscale magnetic fluctuations that arise due to the firehose and mirror instabilities are, therefore,
\begin{subequations}\label{eqn:fhmrgrowth}
\begin{align}
    \frac{\delta B^2_{\rm f}(t)}{\delta B^2_{\rm f}(t_{\rm f})} &\sim \exp\left[ 2 \int^t_{t_{\rm f}} \rmd t' \, \gamma_{\rm f}(t') \right] \sim \exp\left[ \frac{4\Omega_i}{3} \biggl(\frac{\alpha}{\beta_i}\biggr)^{1/4} (|k_\parallel| \vth{i0})^{1/2} \, (t-t_{\rm f})^{3/2} \right], \\*
    \frac{\delta B^2_{\rm m}(t)}{\delta B^2_{\rm m}(t_{\rm m})} &\sim \exp\left[ 2 \int^t_{t_{\rm m}} \rmd t' \, \gamma_{\rm m}(t') \right] \sim \exp\left[ \frac{2\Omega_i}{3} \biggl(\frac{\alpha}{\beta_i}\biggr) (k_\parallel \vth{i0})^2 \, (t-t_{\rm m})^3 \right] ,\label{eqn:growthMR}
\end{align}
\end{subequations}
respectively. This growth in magnetic energy returns the pressure anisotropy to values bounded approximately by the instability thresholds. We argue that, due to the super-exponential growth of the instabilities, this regulation will occur very rapidly once the exponents in \eqref{eqn:fhmrgrowth} become order unity (this neglects order-unity pre-factors and logarithmic corrections). Therefore, the times at which the IAW-driven pressure anisotropy starts to be regulated by the firehose ($t_{\rm f,reg}$) and mirror ($t_{\rm m,reg}$) instabilities should satisfy
\begin{subequations}\label{eqn:tdrain}
\begin{align}
    |k_\parallel| \vth{i0} (t_{\rm f,reg} - t_{\rm f}) \sim (|k_\parallel| \rho_{i0})^{2/3} \biggl(\frac{\beta_i}{\alpha}\biggr)^{1/6} , \label{eqn:tdrainFH}\\*
    |k_\parallel| \vth{i0} (t_{\rm m,reg} - t_{\rm m}) \sim (|k_\parallel| \rho_{i0})^{1/3} \biggl(\frac{\beta_i}{\alpha}\biggr)^{1/3},\label{eqn:tdrainMR}
\end{align}
where we have normalized these times using the IAW wave frequency ${\sim}|k_\parallel|\vth{i0}$.

In order for the regulation times given by \eqref{eqn:tdrain} to be valid, they must satisfy our working assumption that $(t/t_{\rm f}-1)$, $(t/t_{\rm m}-1) \ll 1$. This requires $|k_\parallel|\rho_{i0} \ll \beta^{-1/2}_i (\alpha\beta_i)^{-1/2}$ for the firehose and $|k_\parallel|\rho_{i0} \ll \beta^{-2}_i (\alpha\beta_i)^{-1/2}$ for the mirror. All of the numerical simulations presented in \S\ref{sec:IAWnumerics} easily satisfy the first of these criteria (for the firehose) but not the second (for the mirror). As a result, the onset of the mirror instability occurs late enough that $\Delta(t)$ grows to be ${\gg}1/\beta_i$ before being regulated. In this situation, the Taylor expansion used to obtain \eqref{eqn:growthMR} for the mirror instability, and thus \eqref{eqn:tdrainMR}, is no longer valid. A better approximation is then $\Lambda_{\rm m}(t) \approx \Delta(t) \sim \alpha (k_\parallel \vth{i0}t)^2$, and the argument of the exponential in the amplification factor \eqref{eqn:growthMR} instead reads $(2\Omega_i/5) \alpha^2 (k_\parallel\vth{i0})^4 (t-t_{\rm m})^5$. The corresponding `mirror regulation' time should then satisfy
\begin{equation}\label{eqn:tdrainMR_nonasymp}
    |k_\parallel|\vth{i0} (t_{\rm m,reg}-t_{\rm m}) \sim (|k_\parallel|\rho_{i0})^{1/5} \alpha^{-2/5} .
\end{equation}
\end{subequations}
This prediction should hold so long as $t_{\rm m,reg}$ is less than half of an IAW period, otherwise the oscillating wave would drive the pressure anisotropy negative before the mirror is able to regulate it. Therefore, we require $|k_\parallel|\rho_{i0}\ll\alpha^2$ in order for the prediction \eqref{eqn:tdrainMR_nonasymp} to be valid. The scalings \eqref{eqn:tdrainFH} and \eqref{eqn:tdrainMR_nonasymp} are tested in \S\ref{sec:scaledependence}.

Once the times given by \eqref{eqn:tdrain} are reached, the evolution of the pressure anisotropy will no longer follow \eqref{eqn:IAWdelta}. Instead, it will be regulated to values near the instability thresholds as its unstable portion, {\it viz.}~$\Lambda_{\rm f}(t_{\rm f,reg})$ for firehose and $\Lambda_{\rm m}(t_{\rm m,reg})$ for mirror, is transferred to (mostly transverse) magnetic-field fluctuations, in the case of firehose, and (approximately pressure-balanced) compressive fluctuations, in the case of mirror. Thereafter, these fluctuations will continue to grow in order to maintain locally a marginally unstable plasma, $|\Delta(t)|\sim 2/\beta_i$, as the pressure anisotropy is persistently driven by the oscillating IAW. The result is secular-in-time amplification of the microscale fluctuations \citep{schekochihin08,rosin11,rincon15}. Accompanying this growth is an anomalous `effective collisionality' $\nu_{\rm eff}$, caused by the particles scattering off of these fluctuations. This scattering contributes to maintaining a marginally unstable pressure anisotropy by directly isotropizing the plasma. Once $\nu_{\rm eff}$ attains values comparable to $S\beta_i$, where $S \sim |\rmd\Delta_{\rm IAW}/\rmd t|$ is the rate of pressure-anisotropy production by the oscillating IAW, the enhanced collisionality obviates the need for the microscale fluctuations to continue growing, and the instabilities saturate \citep{kss14}.

This evolution may be formalized somewhat as follows. In the case of the firehose, we adapt the arguments of \citet{schekochihin08} to write the marginally unstable balance between the production of negative pressure anisotropy directly by the IAW, the adiabatic production of positive pressure anisotropy by the growing firehose fluctuations $\delta B_{\rm f}$, and the isotropization of the plasma by the effective collisionality $\nu_{\rm eff}$, heuristically as
\begin{equation}\label{eqn:CGL}
    \D{t}{\Delta} \approx \D{t}{\Delta_{\rm IAW}} + \frac{3}{2} \D{t}{} \frac{\delta B^2_{\rm f}}{B^2_0} - 3\nu_{\rm eff}\Delta \approx 0, ~~{\rm with}~~ \Delta \approx -\frac{2}{\beta_i}. 
\end{equation}
The collisionless secular phase of growth, in which $\nu_{\rm eff}\ll S\beta_i$, is achieved by balancing the first two terms on the right-hand side of \eqref{eqn:CGL}, yielding
\begin{equation}\label{eqn:secular_fh}
    \frac{3}{2} \frac{\delta B^2_{\rm f}(t)}{B^2_0} \approx \frac{3}{2} \frac{\delta B^2_{\rm f}(t_{\rm f,reg})}{B^2_0} + |\Delta_{\rm IAW}(t)| - \frac{2}{\beta_i} .
\end{equation}
This phase of growth is, however, short-lived when $S\beta_i\ll\Omega_i$ \citep{kss14,msk16}. The isotropizing effect of particles' scattering off of the growing firehose fluctuations becomes increasingly important, and ultimately plays the dominant role in regulating the pressure anisotropy:
\begin{equation}\label{eqn:fh_nueff}
    \frac{1}{3\nu_{\rm eff}}\D{t}{\Delta_{\rm IAW}} \approx - \frac{2}{\beta_i} ,
\end{equation}
a sort of Braginskii-like weakly collisional closure. 

In the case of the mirror, the fluctuations grow secularly while trapping an increasing fraction of particles $\xi_{\rm tr} \approx |\delta B_{\rm m}/B_0|^{1/2}$ in the deepening magnetic troughs \citep{schekochihin08,rincon15}. This results in a balance between the production of positive pressure anisotropy by the IAW and what amounts to betatron cooling of the trapped ion  population:
\begin{equation}\label{eqn:secular_mr}
    \D{t}{\Delta} \approx \D{t}{\Delta_{\rm IAW}} - 3 \D{t}{}\left|\frac{\delta B_{\rm m}}{B_0}\right|^{3/2} \approx 0 , ~~{\rm with}~~ \Delta \approx \frac{1}{\beta_i} .
\end{equation}
In principle, this phase would last all the way until $\delta B_{\rm m} / B_0 \approx 0.3$ at $t \gtrsim S^{-1}$, when the ions begin to scatter appreciably off the sharp ends of the mirrors \citep{kss14,riquelme15,msk16}. However, because the large-scale oscillation causes $\Delta_{\rm IAW}$ to change sign halfway through the IAW period, we expect that the mirrors will not have enough time to complete their secular growth to large amplitudes and so the associated strong scattering will not be realized.

%
%
\subsubsection{Effective collisionality and interruption of Landau damping}\label{sec:IAWcoll}

In either situation -- firehose or mirror -- these pressure-anisotropy instabilities will ultimately interfere with phase mixing and collisionless damping by impeding the maintenance of Landau resonances through particle scattering and/or trapping. Accordingly, the rate of damping of the IAW should decrease. 

To gauge just how effective these instabilities are at reducing the Landau damping, let us consider equation \eqref{eqn:fh_nueff} for the balance between pressure-anisotropy production by the IAW and its depletion by the firehose-induced effective collisionality. During this phase, the collisionality is large enough to cause $\Delta_{\rm IAW}$ to depart from the collisionless prediction \eqref{eqn:IAWdelta}. To obtain a rough estimate for the evolution of $\Delta_{\rm IAW}$ during this weakly collisional phase, we adopt the double-adiabatic scaling $\rmd \Delta_{\rm IAW} / \rmd t \approx - 2 (\rmd/\rmd t) (\delta n/n_0) \sim \alpha |k_\parallel|\vth{i0}$. The regulating collisionality must then satisfy
\begin{equation}\label{eqn:nueff}
    \frac{\nu_{\rm eff}}{|k_\parallel|\vth{i0}} \sim \alpha\beta_i
\end{equation}
With $\alpha \gtrsim 2/\beta_i$ being the requirement for the IAW to trigger the firehose instability in the first place, we have $\nu_{\rm eff} \gtrsim |k_\parallel|\vth{i0}$, and so the effective collisionality induced by the firehose is in principle sufficient to halt Landau damping altogether.

%
%
\subsection{Numerical results}\label{sec:IAWnumerics}

We test the ideas put forth in \S\ref{sec:IAWtheory} using hybrid-kinetic simulations with the particle-in-cell code {\tt Pegasus} \citep{kunz14}. The model equations governing the ion distribution function $f_i(t,\bb{r},\bb{v})$ and the electromagnetic fields $\bb{B}(t,\bb{r})$ and $\bb{E}(t,\bb{r})$ are the kinetic Vlasov equation, Faraday's law of induction, and a generalized Ohm's law that assumes quasi-neutrality and includes the inductive and Hall electric fields, as well as the thermo-electric field driven by pressure gradients in a massless electron fluid  (see equations (1)--(4) and (10) in \citealt{kunz14}). Second-order--accurate triangle-shaped stencils are used for interpolating the electromagnetic fields to the particle positions and for depositing moments of $f_i$ onto the grid. A nonlinear $\delta f$ method is used to reduce small-scale noise in the computed ion moments.

All simulations are initialized with $f_i$ given by (\ref{eqn:f0}) on a 2D computational domain that is elongated in the direction of a mean magnetic field $\bb{B}_0 = B_0 \ex$. Our fiducial simulation setup utilizes $N_{\rm ppc} = 10^4$ particles per cell and $3360\times 168$ cells spanning $L_x \times L_y = 1000\rho_{i0} \times 50\rho_{i0}$. This gives a spatial resolution $\Delta x \approx 0.3\rho_{i0}$, the same as was used in prior 2D {\tt Pegasus} simulations of nonlinear firehose and mirror instabilities in shearing plasmas \citep{kss14} and in interrupted shear-Alfv\'{e}n waves \citep{squire17num}. Other simulations with the same values of $N_{\rm ppc}$, $L_y/\rho_{i0}$, and $\Delta x/\rho_{i0}$ but with $L_x/\rho_{i0}=2000$, $4000$, $8000$, $16000$, and $32000$ have also been performed, yielding qualitatively similar results (see figures~\ref{fig:scan} and \ref{fig:nuscan}).\footnote{Those with $L_x \le 500\rho_{i0}$ show non-asymptotic behaviour, as the mirror modes do not have enough time to grow before the density perturbation reverses its oscillation. The $L_x/\rho_{i0}=8000$, $16000$, and $32000$ simulations were run only long enough to observe the triggering of firehose and mirror instabilities and their saturation, in support of figure~\ref{fig:scan}. For context, the cost for one of these simulations to run for one IAW period is ${\approx}2.6\times 10^6 (L_x/32000\rho_{i0})^2$ CPU-hrs.} The electrons are assumed to be isothermal with temperature $T_e = T_{i0}$. To maintain contact with the calculations of wave interruption by \citet{squire17num}, we focus primarily on results using a $\betaio = 100$ plasma (although other values of $\betaio$ have also been investigated). The initial amplitude of the IAW, $\alpha$, is varied above and below $2/\betaio$. In what follows, $\langle\,\cdot\,\rangle$ denotes a spatial average over all cells, while $\langle\,\cdot\,\rangle_y$ denotes a spatial average only over the $y$ direction.

%
%
\subsubsection{Generation of pressure anisotropy and triggering of kinetic instabilities}

%
%
\begin{figure}
    \centering
    \includegraphics[width=0.48\textwidth]{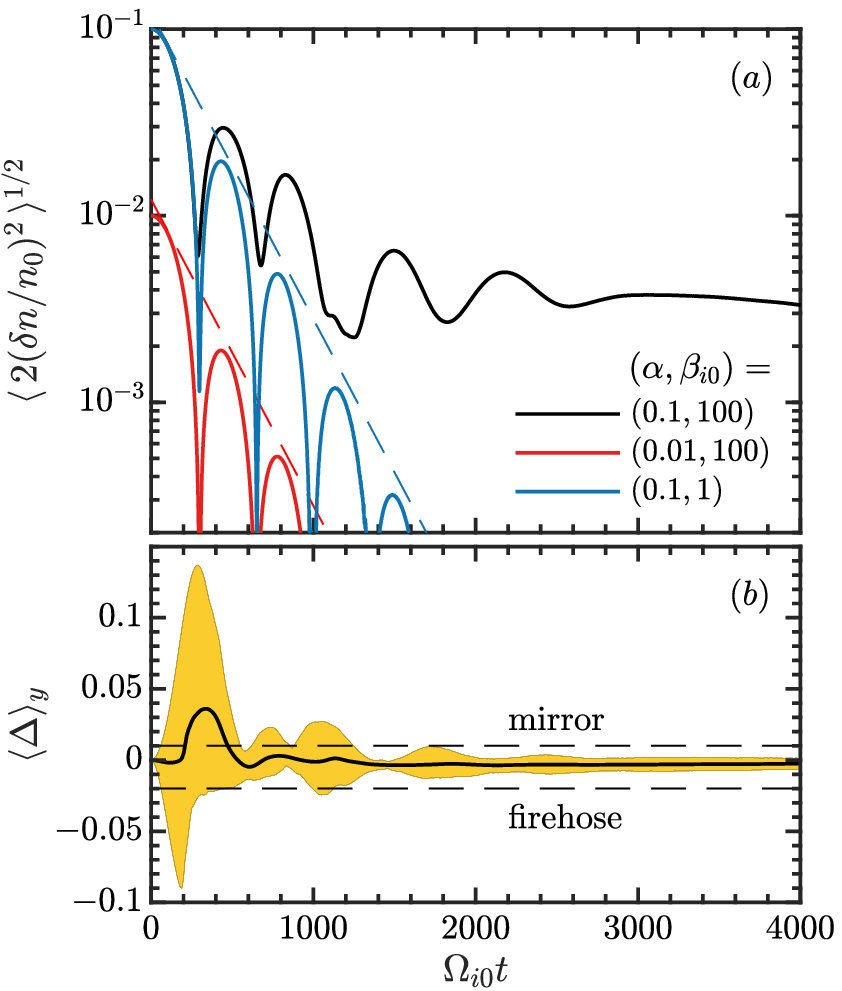}
    \includegraphics[width=0.48\textwidth]{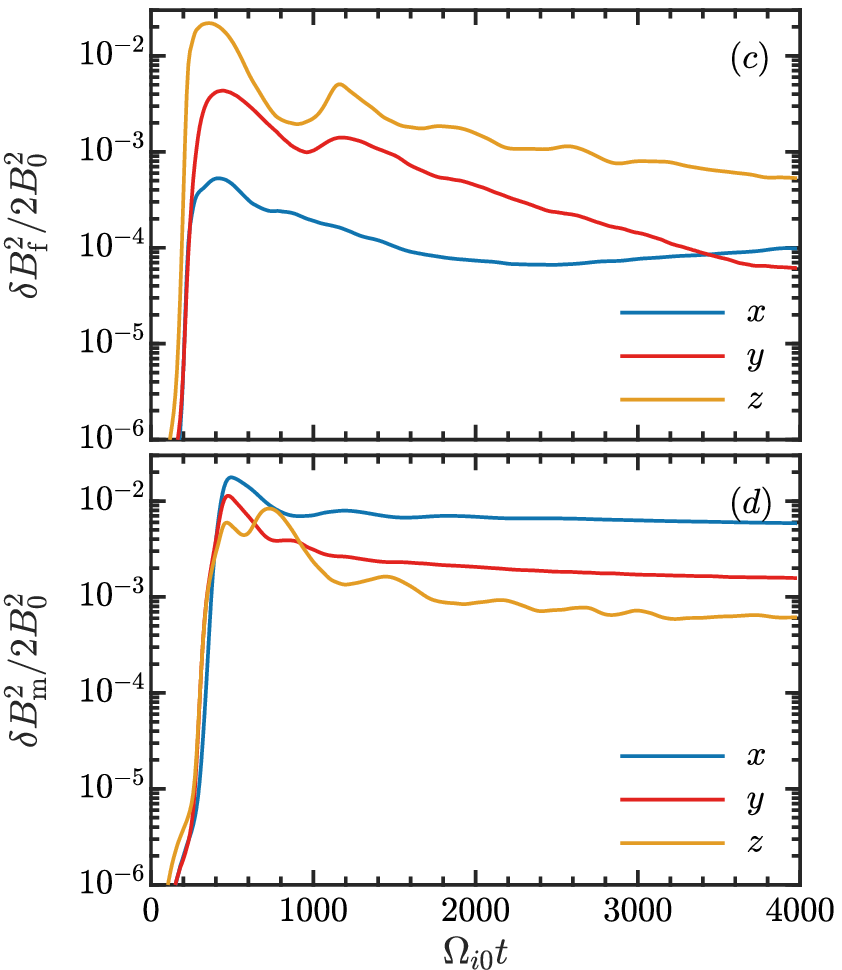}
    \caption{Interrupted collisionless damping of an IAW of wavelength $\lambda_\parallel = 1000\rho_{i0}$, initial frequency $\omega = 1.45|k_\parallel|\vth{i0}$, and initial amplitude $\alpha=0.1$. ($a$) Amplitude of density perturbation; dashed line shows theoretical Landau damping at the predicted linear rate $\gamma/|k_\parallel|\vth{i0} = 0.60$ (see \eqref{eqn:IAW}). For comparison, identical IAWs but with $\alpha=0.01$ (red line) or with $\betaio=1$ (blue line) are shown to be damped at the Landau rate. ($b$) Mean (black line) and spread in $x$ (yellow region) of the $y$-averaged pressure anisotropy, $\langle\Delta \doteq p_\perp/p_\parallel - 1\rangle_y$, vs.~time. Mirror (${\simeq}1/\beta_i$) and firehose (${\simeq}-2/\beta_i$) instability thresholds are denoted by the dashed lines. The energy densities of the $x$, $y$, and $z$ components of the magnetic-field fluctuations associated with the firehose ($\delta B^2_{\rm f}/2B^2_0$) and mirror ($\delta B^2_{\rm m}/2B^2_0$) instabilities are shown in panels ($c$) and ($d$), respectively (see text for their definitions).}
    \label{fig:iaw_damping}
\end{figure}

Figure~\ref{fig:iaw_damping}({\it a}) shows the evolution of $\langle 2(\delta n/n_0)^2\rangle^{1/2}$ using our fiducial setup for $\alpha = 0.1$ (black) and $\alpha = 0.01$ (red), compared to their expected exponential decay by linear Landau damping (dashed lines; (\ref{eqn:IAW}) gives $\gamma/|k_\parallel|\vth{i0} = 0.60$ for $T_e/T_i=1$). While the $\alpha=0.01$ perturbation is efficiently Landau damped, the damping of the $\alpha=0.1$ ($>2/\betaio$) perturbation slows down and ultimately stalls. An IAW with $\alpha=0.1$ and $\betaio=1$ (blue) is also shown to be Landau damped to small amplitude.\footnote{The IAWs in the $(\alpha,\betaio)=(0.01,100)$ and $(0.1,1)$ runs are Landau-damped all the way down to amplitudes comparable to the $N_{\rm ppc}$-dependent noise floor (here, ${\sim}10^{-4}$). The value of $\langle 2(\delta n/n_0)^2\rangle^{1/2}$ in the $(\alpha,\betaio)=(0.1,100)$ IAW (whose decay is stalled) includes a ${\sim}10\%$ contribution at late times from small-scale density fluctuations caused by the mirror instability.} Figure~\ref{fig:iaw_damping}({\it b}) displays the spread (i.e., minimum to maximum) in $x$ (yellow region) and mean (black line) of the $y$-averaged pressure anisotropy $\langle\Delta\rangle_y$ vs.~time for this stalled-decay perturbation. For $\Omega_{i0} t \lesssim 300$ (roughly half a wave period), the extrema of $\langle\Delta\rangle_y$ follow the theoretical prediction (\ref{eqn:IAWdelta}). Thereafter, $|\Delta|$ is large enough for the growth of firehose and mirror instabilities and the consequent relaxation of the pressure anisotropy to overwhelm the production of $\Delta$ (with firehose occurring earlier than mirror, as predicted). Figures~\ref{fig:iaw_damping}({\it c,d}) show the energy densities of the $x$, $y$, and $z$ components of the magnetic-field fluctuations associated with the firehose and mirror instabilities, defined by the spatial regions in which they primarily occur: $\delta B^2_{\rm f} \doteq \langle\delta B^2_{x,y,z}\rangle_{250 \le x/\rho_{i0} \le 750}$ for firehose, and  $\delta B^2_{\rm m} \doteq \langle\delta B^2_{x,y,z}\rangle_{0\le x/\rho_{i0}<250 \,\cup\, 750<x/\rho_{i0}\le1000}$ for mirror. As we demonstrate below (see figure~\ref{fig:iaw_nueff}), these fluctuations locally scatter and trap particles, respectively, inhibiting strong Landau resonances and thereby stalling the Landau damping of the IAW.

%
%
\begin{figure*}
\centering
\includegraphics[width=\textwidth]{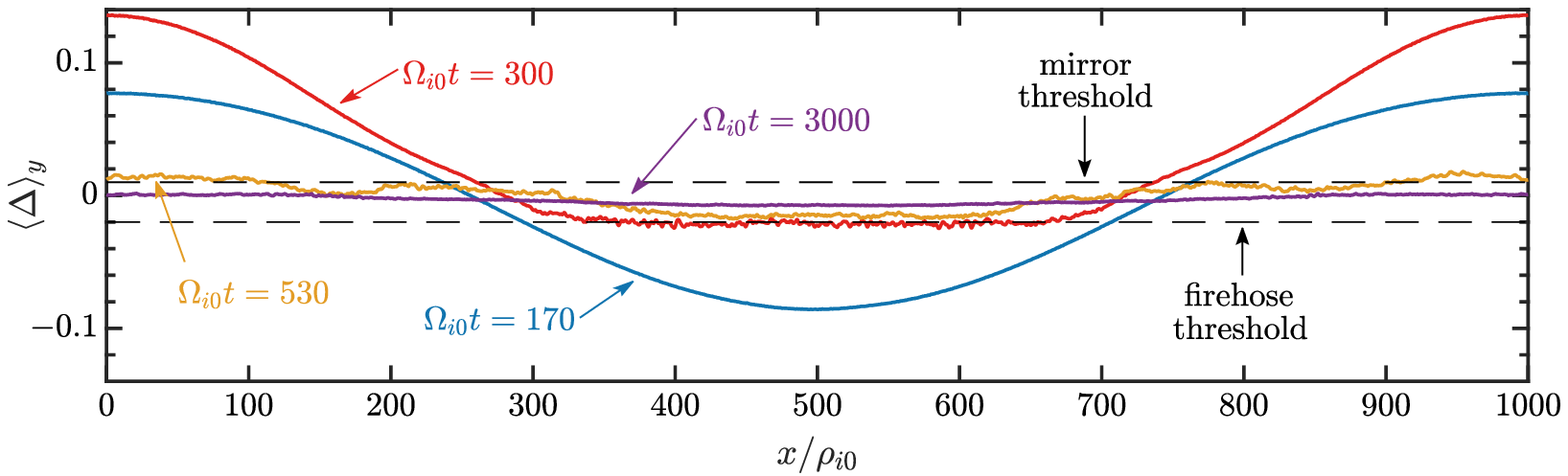} \\
\includegraphics[width=\textwidth]{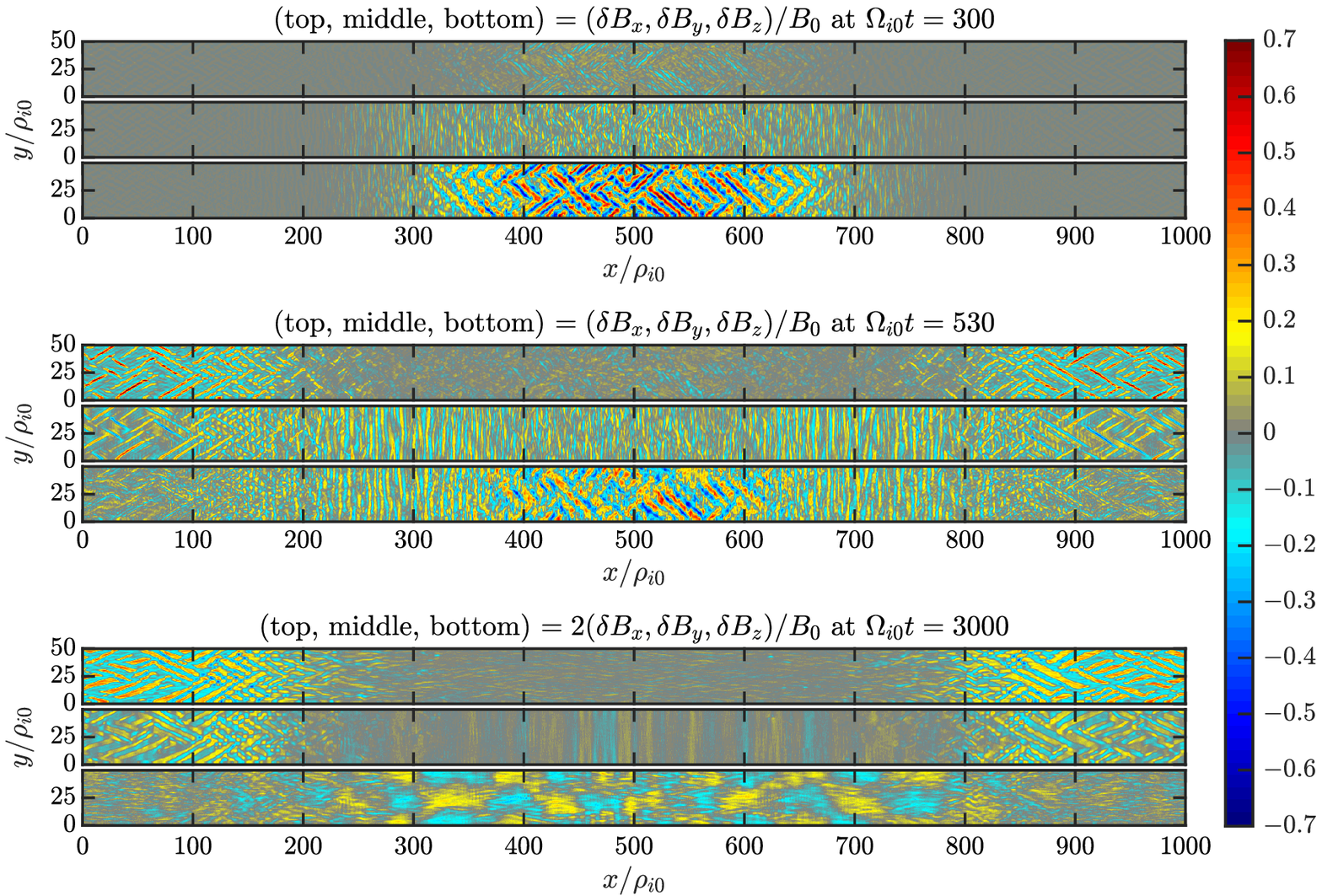}
\caption{(top panel) Spatial profile of $y$-averaged pressure anisotropy at $\Omega_{i0} t = 170$ (when $\langle\Delta\rangle_y$ is most negative), $300$ (at firehose regulation), $530$ (at mirror regulation), and $3000$ (when $\langle\nu_{\rm eff}\rangle \lesssim 10^{-3}\Omega_{i0}$ -- see figure~\ref{fig:iaw_nueff}). (bottom panels) Spatial profiles of magnetic-field fluctuations (normalized to $B_0$) at $\Omega_{i0} t = 300$, $530$, and $3000$. (The fluctuations shown in the $\Omega_{i0}t = 3000$ panels are multiplied by $2$ to enhance their visibility.) All plots correspond to the $(\alpha,\betaio)=(0.1,100)$ IAW featured in figure~\ref{fig:iaw_damping}.}
\label{fig:iaw_dB}
\end{figure*}

The production of unstable magnetic-field fluctuations and consequent regulation of the pressure anisotropy is also illustrated in figure~\ref{fig:iaw_dB}, which shows $\langle\Delta\rangle_y$ and $\delta B_{x,y,z}$ as functions of $x$ at various times. At $\Omega_{i0} t = 170$ (blue line), the pressure anisotropy attains its global minimum (${\simeq}-0.09$), after which the $\langle\Delta\rangle_y < 0$ region (within $250 \lesssim x/\rho_{i0} \lesssim 750$) is quickly (by $\Omega_{i0} t = 300$) regulated to the firehose threshold (red line). The magnetic-field fluctuations in this region are oriented primarily in the $z$ direction and are oblique with respect to the mean magnetic field. By $\Omega_{i0} t = 530$ (yellow line), the regions with $\langle\Delta\rangle_y > 0$ ($0\lesssim x/\rho_{i0}\lesssim 250$ and $750 \lesssim x/\rho_{i0} \lesssim 1000$) are well regulated to the mirror threshold. Indeed, the magnetic-field fluctuations in these regions are mirror-mode polarized, with $\delta B_\perp / \delta B_\parallel \simeq 0.36 \simeq {\rm max}(\Delta - 1/\beta_i)^{1/2}$ \citep{hellinger07}. At late times (e.g., $\Omega_{i0} t = 3000$, purple line), the plasma is nearly pressure-isotropic. While the mirror modes appear to be long-lived, pressure-balanced structures, the firehose modes become stably propagating Alfv\'{e}n waves with parallel wavenumbers comparable to $\rho^{-1}_i$ (as also seen in the simulations of Alfv\'{e}n-wave interruption by \citealt{squire17num}).

%
%
\subsubsection{Effective collisionality: particle scattering and trapping}

%
%
\begin{figure*}
\centering
\includegraphics[width=\textwidth]{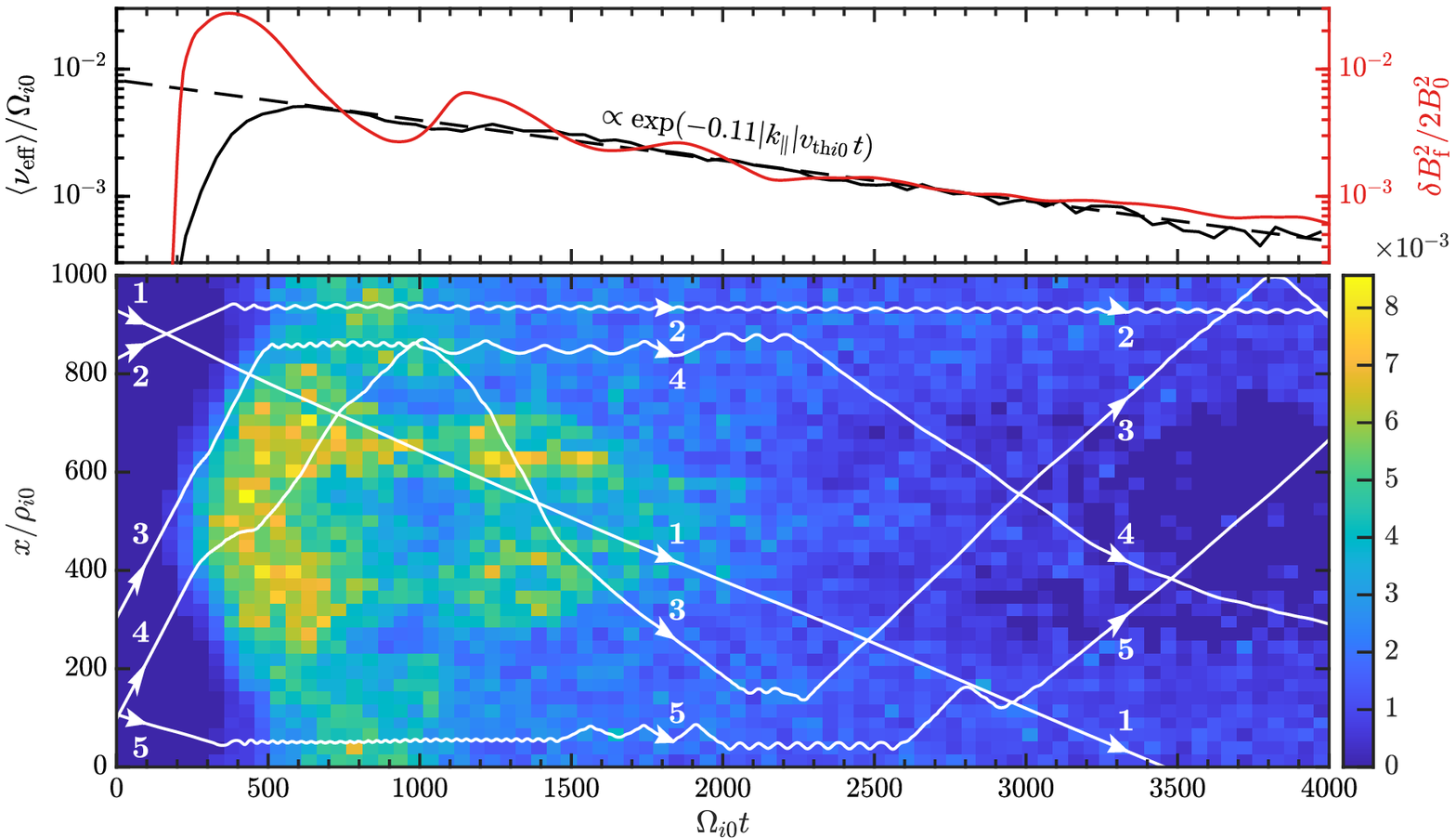} \\
\includegraphics[width=\textwidth]{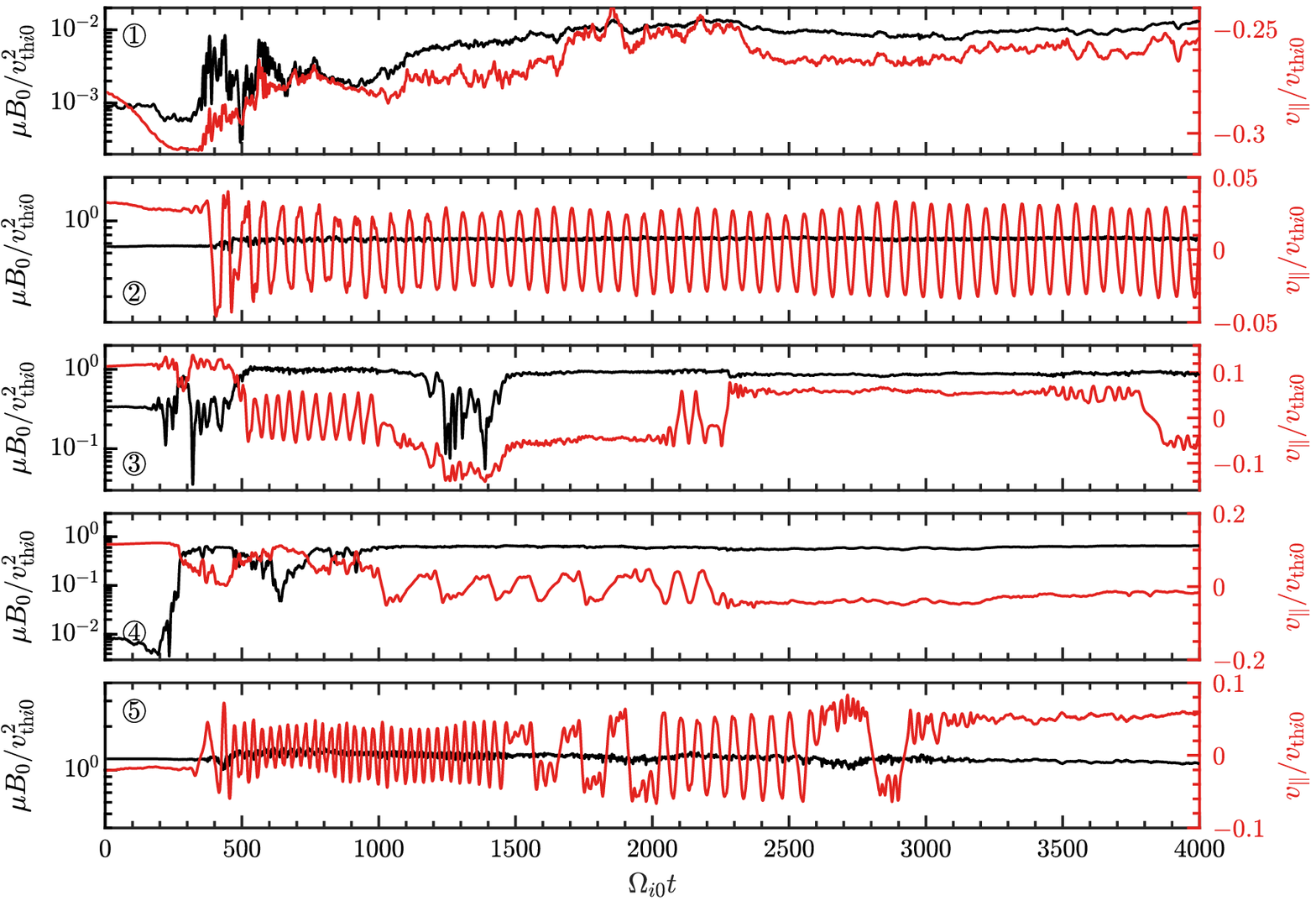}
\caption{(top panel) Box-averaged effective collision frequency, $\langle\nu_{\rm eff}\rangle/\Omega_{i0}$ (black), and magnetic energy of firehose fluctuations, $\delta B^2_{\rm f}/2B^2_0$ (red), vs.~time. (middle panel) Space-time diagram of $\langle\nu_{\rm eff}\rangle_y/\Omega_{i0}$ (colour); overlaid white lines are trajectories of five representative particles. (bottom panels) Magnetic moments $\mu(t)$ (black) and parallel velocities $v_\parallel(t)$ (red) of those particles, labelled \circleit{1}--\circleit{5}. All plots correspond to the $(\alpha,\betaio)=(0.1,100)$ IAW featured in figures~\ref{fig:iaw_damping} and \ref{fig:iaw_dB}.}
\label{fig:iaw_nueff}
\end{figure*}

Figure~\ref{fig:iaw_nueff} provides the link between these magnetic-field fluctuations and the evolution shown in figure~\ref{fig:iaw_damping}. The top two panels illustrate the effective collisionality of the plasma $\nu_{\rm eff}$, computed following the method used in \citet{kss14}, \citet{msk16}, and \citet{squire17num}.\footnote{Namely, the mean time $\tau$ for the particle adiabatic invariant $\mu=v^2_\perp/2B$ to change by a factor of $\rme$ is calculated using 6,720 tracked particles; then $\nu_{\rm eff} \doteq 1/\tau$. In practice, factor-$\rme$ changes are rare, and so we instead measure $\tau_\kappa$ for factor-$\kappa$ changes and set $\nu_{\rm eff}\doteq(\ln\kappa)^2/\tau_\kappa$. We chose $\kappa=1.2$, but checked that other values $1.1\lesssim \kappa\lesssim 2$ gave similar results. That this choice of definition of $\nu_{\rm eff}$ is a good one quantitatively as well as qualitatively is supported by figure~\ref{fig:iaw_q}.} The box-averaged collisionality $\langle\nu_{\rm eff}\rangle$ (black line) increases dramatically once the firehose modes are triggered, peaking approximately when the IAW stops Landau damping at a value $\nu_{\rm eff} \sim |k_\parallel|\vth{i0}$. Thereafter, its steady exponential decrease follows the decaying magnetic energy of the firehose fluctuations (red line), {\it viz.}, $\langle\nu_{\rm eff}\rangle/\Omega_{i0} \approx \delta B^2_{\rm f}/2B^2_0$, as it should if this scattering is due to pitch-angle scattering of particles off ion-Larmor-scale magnetic fluctuations \citep[e.g.,][]{kp69,schlickeiser89}. What sets the exponential decay rate of the firehose fluctuations, and thus the anomalous collisionality, is not yet fully understood, but a scan across $L_x/\rho_{i0}=1000$--$16000$ returned the empirical result that their decay rate is ${\approx}0.1 |k_\parallel|\vth{i0}$ with $|k_\parallel|=2\upi/L_x$ (see figure~\ref{fig:nuscan}($b$)). If this trend holds at even larger scale separations, it would imply that the observed interruption of Landau damping by scattering off firehose fluctuations is an asymptotically valid effect. Importantly, {\em if IAWs were to be continuously replenished in a magnetized, high-$\beta$ plasma on a timescale comparable to or smaller than their oscillation period, the resulting dynamics would never be collisionless.} We revisit this idea in \S\ref{sec:IAWLangevin}.

The space-time evolution of $\nu_{\rm eff}/\Omega_{i0}$ is also shown in figure~\ref{fig:iaw_nueff} (colour), with tracks of five representative particles overlaid (white lines, labelled $1$--$5$). Those particles' adiabatic invariants ($\mu$; black lines) and parallel velocities ($v_\parallel$; red lines) are plotted in the bottom five panels, correspondingly labelled \circleit{1}--\circleit{5}. All particles begin their lives by streaming freely (i.e., $x(t) = x(0) + v_\parallel t$) and thus phase-mixing the distribution function. Between $\Omega_{i0} t \approx 200$--$300$, particles 1, 3, and 4 encounter firehose fluctuations and promptly scatter, changing their $\mu$ by large factors. In contrast, particles 2 and 5 first encounter mirror modes at $\Omega_{i0} t \approx 400$, become trapped, and bounce, conserving $\mu$ while exhibiting oscillations in $v_\parallel$ about $0$. After this initial scattering or trapping, these particles either stay trapped in mirrors (\circleit{2}), possibly leaking out at late times (\circleit{5}); or eventually encounter a mirror and become trapped, only subsequently to scatter out of the mirror and once again encounter $\mu$-breaking firehose fluctuations (\circleit{3}, \circleit{4}); or continue scattering while streaming un-trapped, but with decreasing changes in $\mu$ as the fluctuations decay (\circleit{1}). Note that particle 3's $\mu$ conservation is broken at $\Omega_{i0} t \approx 1200$ when the particle encounters a second burst of firehose fluctuations produced by the oscillating IAW.

%
%
\subsubsection{Reduction of heat conduction by firehose and mirror}

\begin{figure*}
    \centering
    \includegraphics[width=\textwidth]{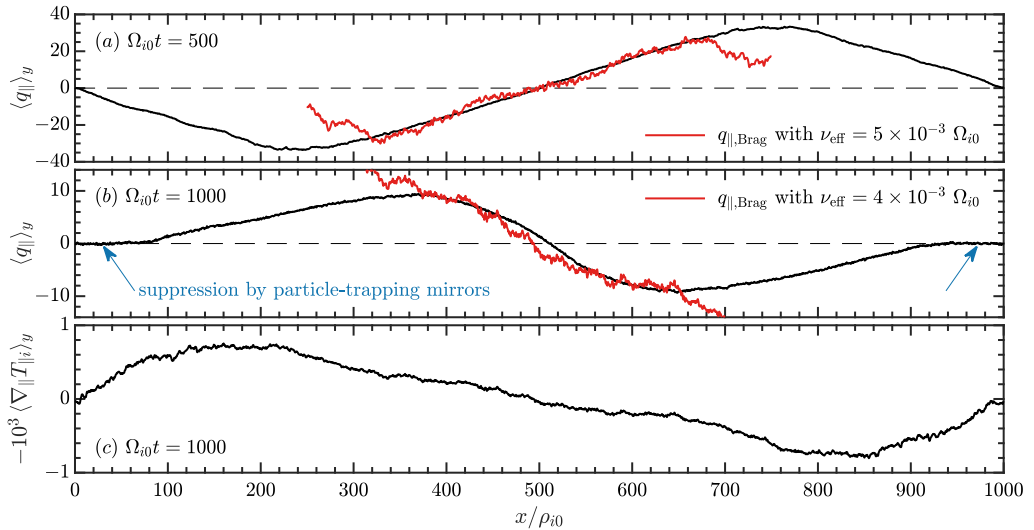}
    \caption{Spatial profile of $y$-averaged parallel ion heat flux $\langle q_\parallel \rangle_y$ at ($a$) $\Omega_{i0}t = 500$ (near the time when the firehose-induced collisionality is at its peak value; see figure~\ref{fig:iaw_nueff}, top panel) and ($b$) $\Omega_{i0}t = 1000$ (after mirror-mode saturation). In ($b$), near-complete suppression of $\langle q_\parallel \rangle_y$ by particle trapping in mirrors occurs (indicated by the blue arrows), despite the presence of a strong ion temperature gradient (panel ($c$)). The heat flux in the central (firehose-unstable) region is accurately described by a Braginskii heat flux (red) using the effective collisionality measured directly from tracked particles (see (\ref{eqn:bragQ})).}
    \label{fig:iaw_q}
\end{figure*}

An enhanced effective collisionality in the firehose-unstable regions and efficient particle trapping in the mirror-unstable regions suggest that appreciable reductions in the (ion) heat conductivity occur throughout the IAW. In regions where the firehose fluctuations break $\mu$ conservation at the rate $\nu_{\rm eff}$, we may anticipate the parallel flow of parallel heat
\begin{equation}\label{eqn:bragQ}
q_{\parallel} \doteq \int\rmd^3\bb{v} \, m_i (v_\parallel-u_{\parallel i})^3 f_i \approx -\frac{3}{2} n \frac{v^2_{{\rm th}\parallel i}}{\nu_{\rm eff}} \nabla_\parallel T_{\parallel i} \doteq q_{\parallel,{\rm Brag}} ,
\end{equation}
following \citet{braginskii65} and using a pitch-angle-scattering (Lorentz) collision operator with collision frequency $\nu_{\rm eff}$; $u_{\parallel i} \doteq (1/n)\int\rmd^3\bb{v}\, v_\parallel f_i$ is the parallel ion bulk velocity. In regions where the mirror fluctuations are in their late nonlinear stage of evolution and efficiently trap a majority fraction of the nearby particles, $q_\parallel$ should be close to $0$. Figure~\ref{fig:iaw_q} confirms these expectations. 
%
%
Once the firehose fluctuations become efficient scatterers ($\Omega_{i0}t\gtrsim{500}$; panel ($a$)), the Braginskii heat flux $q_{\parallel,{\rm Brag}}$ given by \eqref{eqn:bragQ} (red line), with $\nu_{\rm eff}$ taken from figure~\ref{fig:iaw_nueff}, is an excellent description of the actual heat flux (black line) in the firehose-infested region. And once the mirrors become efficient particle traps ($\Omega_{i0}t\gtrsim{1000}$; figure~\ref{fig:iaw_q}($b$)), $q_\parallel$ is substantially suppressed in those regions (marked by the blue arrows) -- the conductivity there is effectively zero, despite the persistent temperature gradient (figure~\ref{fig:iaw_q}($c$)). A rough quantitative measure of the suppression factor may be obtained by calculating $\langle q_\parallel\rangle_y / \langle -\nabla_\parallel T_{\parallel i}\rangle_y$ in the mirror-infested regions of the simulation and comparing it to the corresponding collisionless value without mirrors obtained using a linear Landau-fluid model, {\it viz.}, $(2/\sqrt{\upi}) n_0 \vth{i0} / |k_\parallel|$ (see equation (39) of \citealt{shd97}). The result gives a suppression factor ${\sim}200$. (A suppression factor of ${\sim}5$ was calculated by \citealt{komarov16} for the {\em electron} parallel heat conductivity in the presence of saturated, ion-Larmor-scale mirrors.)

%
%
\subsubsection{Dependence on scale separation}\label{sec:scaledependence}

%
%
\begin{figure}
    \centering
    \includegraphics[width=\textwidth]{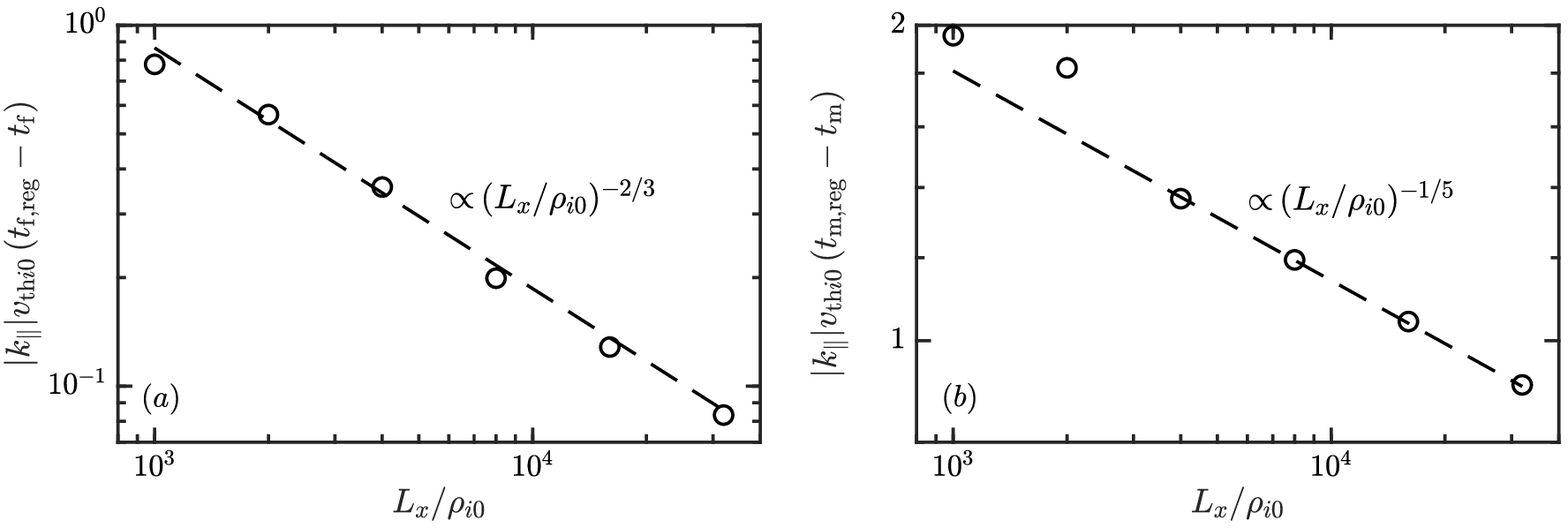}
    \newline
    \includegraphics[width=\textwidth]{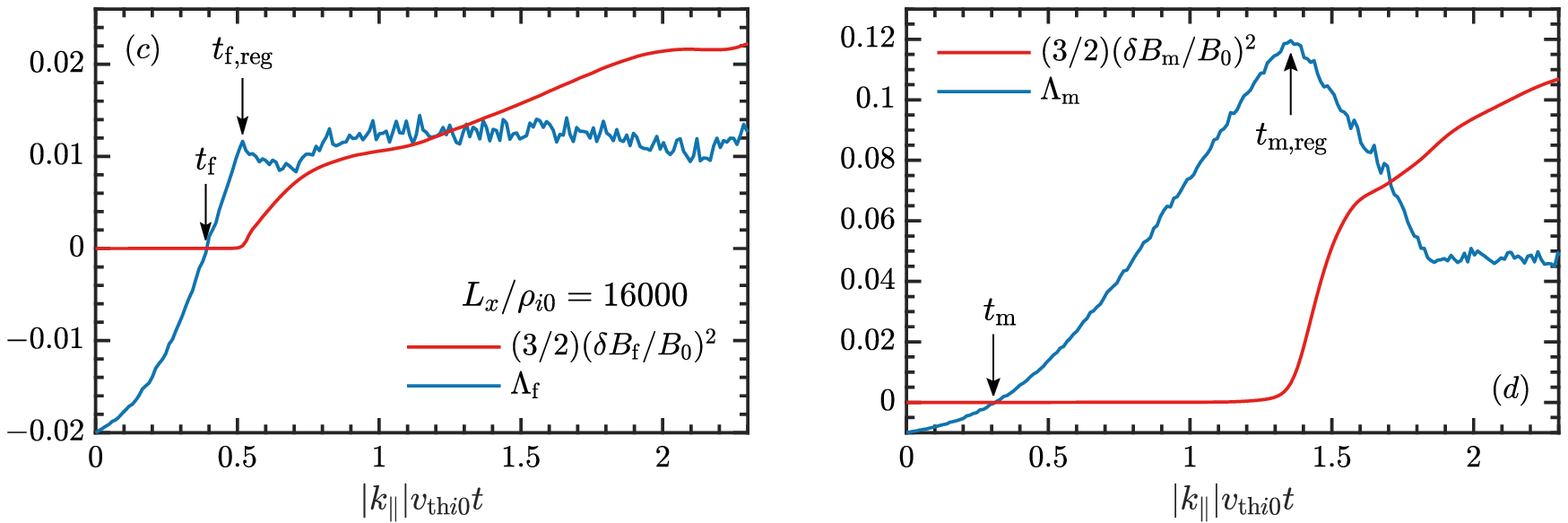}
    \newline
    \includegraphics[width=\textwidth]{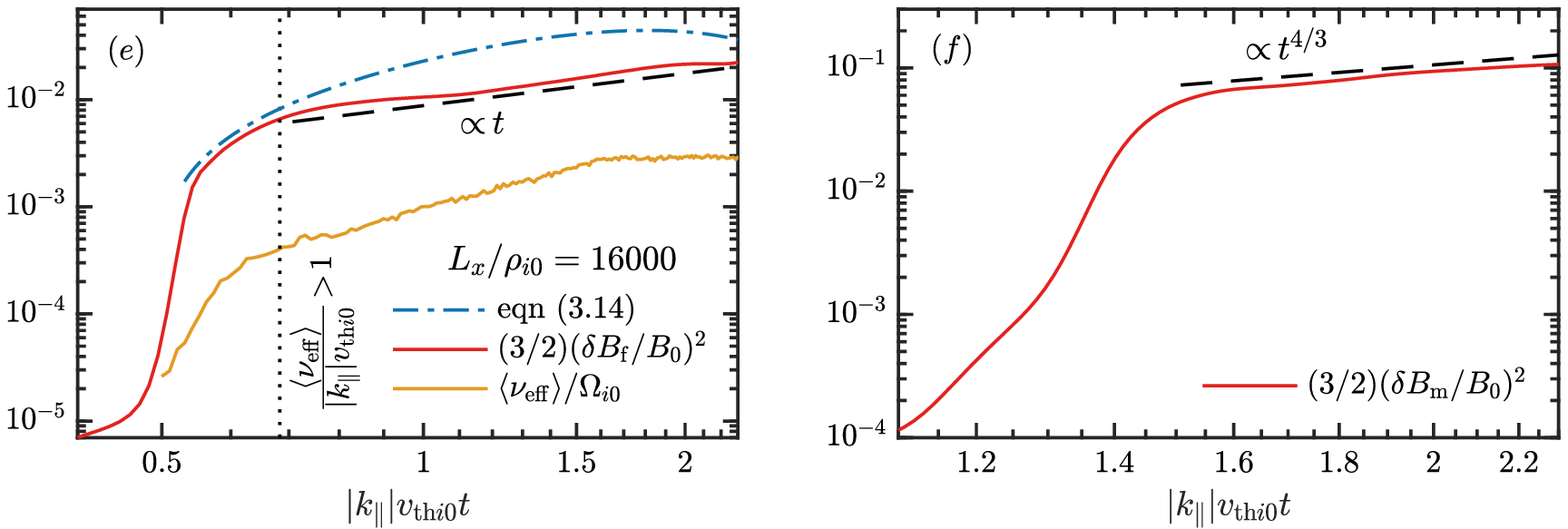}
    \caption{($a$) Time elapsed between $t_{\rm f}$, when the plasma first becomes firehose unstable, and $t_{\rm f,reg}$, when the negative pressure anisotropy is rapidly regulated by the firehose instability, vs.~scale separation $L_x/\rho_{i0}$. The dashed line is the theoretical prediction, \eqref{eqn:tdrainFH}. ($b$) As in ($a$) but for positive pressure anisotropy and the mirror instability; the dashed line is the theoretical prediction  \eqref{eqn:tdrainMR_nonasymp}. ($c$) Early evolution of the firehose instability parameter $\Lambda_{\rm f}$ (blue) and (three times) the energy density of the firehose magnetic-field fluctuations (red) for $L_x/\rho_{i0}=16000$. The times $t_{\rm f}$ and $t_{\rm f,reg}$ are indicated. ($d$) As in ($c$) but for the mirror instability. Also for $L_x/\rho_{i0}=16000$: ($e$) logarithmic evolution of the energy density of the firehose magnetic-field fluctuations (red) and of the box-averaged effective collision frequency (orange); and ($f$) logarithmic evolution of the energy density of the mirror magnetic-field fluctuations (red). These final panels highlight the secular phase of unstable growth, occurring with power laws close to ${\propto}t$ and ${\propto}t^{4/3}$ after the plasma becomes weakly collisional, $\langle\nu_{\rm eff}\rangle/|k_\parallel|\vth{i0}>1$ (vertical dotted line). Before that time, $(3/2)(\delta B_{\rm f}/B_0)^2$ follows well the collisionless prediction \eqref{eqn:secular_fh} (blue dot-dashed line). All times are normalized using $|k_\parallel|\vth{i0}=2\upi\Omega_{i0}(\rho_{i0}/L_x)$; all data refer to the case  $(\alpha,\betaio)=(0.1,100)$.}
    \label{fig:scan}
\end{figure}

As a last point of comparison with the theory developed in \S\ref{sec:IAWtheory}, we demonstrate how some of our quantitative results depend on scale separation, $L_x/\rho_{i0}$. First, in figure~\ref{fig:scan}($a$), we show that the values of $|k_\parallel|\vth{i0}(t_{\rm f,reg}-t_{\rm f})$ measured in IAW simulations with $(\alpha,\betaio)=(0.1,100)$ across a range of $L_x/\rho_{i0}$ are in remarkable agreement with the predicted scaling \eqref{eqn:tdrainFH}; the omitted proportionality constant in \eqref{eqn:tdrainFH} is found empirically to be ${\approx}8$. For the mirror instability, we find $|k_\parallel|\vth{i0}(t_{\rm m,reg}-t_{\rm m})\propto(L_x/\rho_{i0})^{-1/5}$ once $L_x/\rho_{i0} \ge 4000$, consistent with the corresponding prediction \eqref{eqn:tdrainMR_nonasymp}, with a proportionality constant ${\approx}2$. This is as it should be, because the predicted asymptotic scaling \eqref{eqn:tdrainMR} is not applicable unless $|k_\parallel|\rho_i\ll \betaio^{-2} (\alpha\betaio)^{1/2}$, a condition that is never satisfied even in our very largest numerical simulations. These scalings demonstrate that, in the limit $\rho_{i0}/L_x\rightarrow 0$, the pressure anisotropy will be regulated by the kinetic instabilities almost instantly after the system becomes unstable at $t\sim t_{\rm f}, t_{\rm m}$ (see \eqref{eqn:tFHMR}).\footnote{A similar conclusion was reached by \citet{kss14} in their numerical study of shear-driven firehose and mirror instabilities -- see their figures 1($b$) and 6($b$).}

In panels ($c$) and ($d$), we show the early-time evolution of the firehose ($\Lambda_{\rm f}$) and mirror ($\Lambda_{\rm m}$) instability parameters (blue), as well as of (three times) the energy densities of the magnetic-field fluctuations associated with the firehose and mirror instabilities (red), all from the IAW simulation with $(\alpha,\betaio)=(0.1,100)$ and $L_x/\rho_{i0}=16000$.\footnote{The factor of 3 is included to make contact with \eqref{eqn:CGL}. Indeed, $(3/2)(\delta B_f/B_0)^2$ at the start of its secular phase of growth is comparable to ${\rm max}(\Lambda_{\rm f})$, consistent with the discussion in \S\ref{sec:IAWtheory_inst}.} The times at which the plasma goes firehose (mirror) unstable, $t_{\rm f}$ ($t_{\rm m}$), and at which the pressure anisotropy is regulated by the growth of magnetic-field fluctuations, $t_{\rm f,reg}$ ($t_{\rm m,reg}$), are indicated by the arrows. The observed behaviour matches that predicted in \S\ref{sec:IAWtheory_inst}: at $t \sim t_{\rm reg}$, rapid super-exponential growth of the firehose and mirror fluctuations curtails the production of pressure anisotropy by the IAW, after which the fluctuations grow secularly to help maintain a marginally unstable pressure anisotropy. This secular growth is highlighted in panels ($e$) and ($f$) using logarithmic axes; panel ($e$) also includes the evolution of the box-averaged effective collision frequency (orange line). For times during which $\langle\nu_{\rm eff}\rangle < |k_\parallel|\vth{i0}$, the secular growth of the firehose fluctuations follows well the collisionless prediction \eqref{eqn:secular_fh} (blue dot-dashed line). Once $\langle\nu_{\rm eff}\rangle$ becomes large enough to interfere with the collisionless evolution (i.e., for times to the right of the vertical dotted line at $|k_\parallel|\vth{i0} \approx 0.7$), the firehose fluctuations continue to grow secularly but with a shallower slope close to $\delta B^2_{\rm f} \propto t$. The mirror fluctuations satisfy $\delta B^2_{\rm m} \propto t^{4/3}$ during their secular phase. These power laws match those measured by \citet{kunz14} during the secular growth phase of their dedicated simulations of driven firehose and mirror instabilities, and are consistent with \eqref{eqn:secular_fh} and \eqref{eqn:secular_mr} after the effects of suppressed phase mixing and Landau damping are accounted for in modifying the drive, $\rmd \Delta_{\rm IAW}/\rmd t$. The consequent regulation of the pressure anisotropy is aided by an increasing $\langle\nu_{\rm eff}\rangle$, which attains a maximum value ${\approx}0.003\Omega_{i0} \approx 8|k_\parallel|\vth{i0}$ at $|k_\parallel|\vth{i0} \sim 2$, when the oscillating IAW changes the sign of $\rmd\Delta_{\rm IAW}/\rmd t$ and the growth of the firehose fluctuations is disabled.

%
%
\begin{figure}
    \centering
    \includegraphics[width=\textwidth]{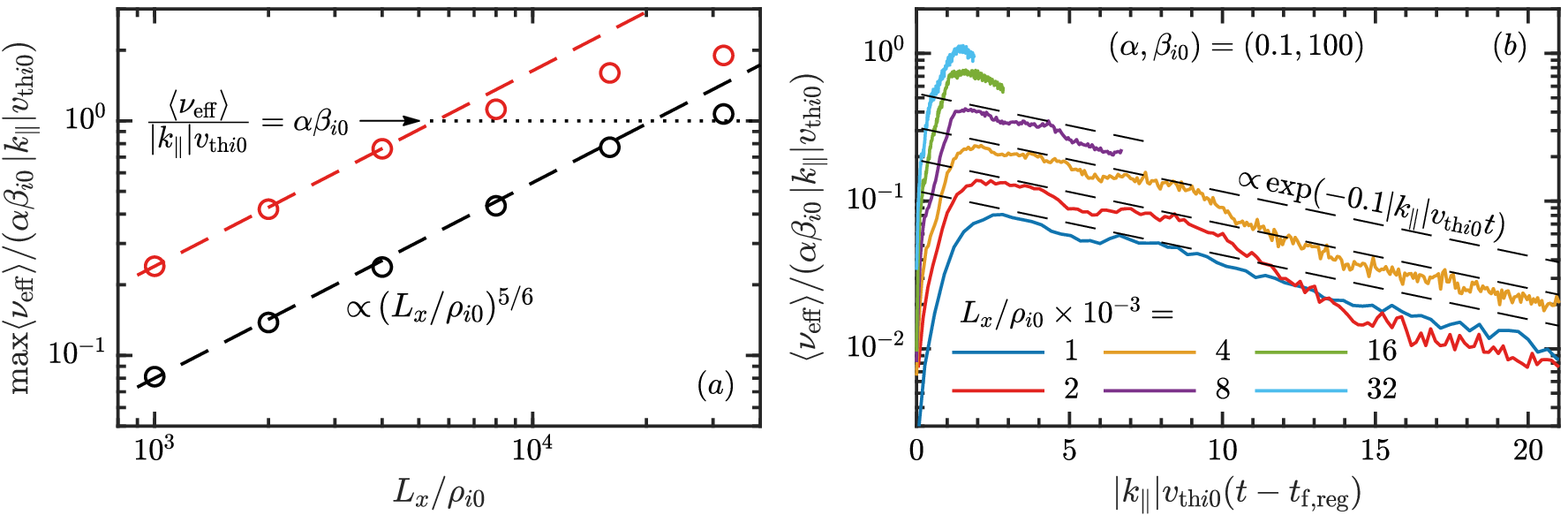}
    \newline
    \includegraphics[width=0.67\textwidth]{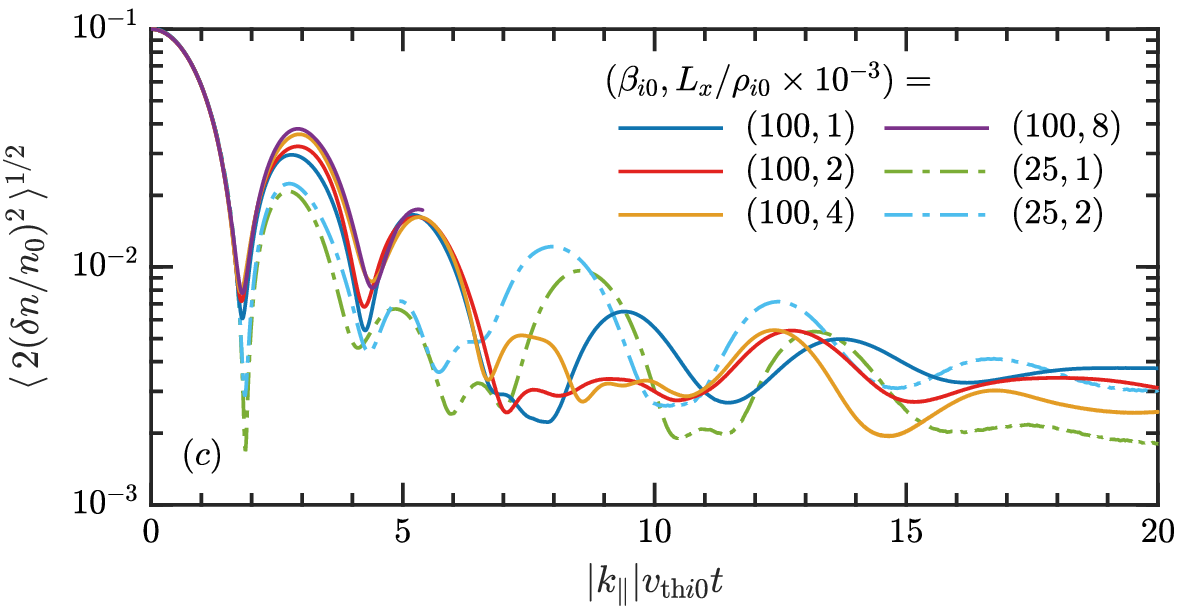}
    \caption{($a$) Maximum value of the box-averaged effective collision frequency $\langle\nu_{\rm eff}\rangle$, normalized using $\alpha\betaio|k_\parallel|\vth{i0}$ with $|k_\parallel|=2\upi/L_x$. Its value increases with scale separation as $(L_x/\rho_{i0})^{5/6}$ until a value ${\sim}1$ (dotted line; see \eqref{eqn:nueff}) is achieved. Black (red) circles correspond to $\betaio=100$ ($\betaio=25$). ($b$) Time evolution of $\langle\nu_{\rm eff}\rangle$ for different $L_x/\rho_{i0}$ at $\betaio=100$. In all cases with long enough integration times, $\langle\nu_{\rm eff}\rangle$ is measured to decay exponentially at a rate ${\approx}0.1|k_\parallel|\vth{i0}$, independent of scale separation. ($c$) Root-mean-square density fluctuation vs.~time for various combinations of $\betaio$ and $L_x/\rho_{i0}$. For all points and curves, $\alpha=0.1$.}
    \label{fig:nuscan}
\end{figure}

Figure~\ref{fig:nuscan} shows how the characteristics of $\langle\nu_{\rm eff}\rangle$ depend on $L_x/\rho_{i0}$, using the same parameter scan as in figure~\ref{fig:scan}($a$) plus an additional scan with $(\alpha,\betaio)=(0.1,25)$. Panel ($a$) confirms that the maximum effective collision frequency in all of our runs is large enough to inhibit Landau damping (i.e., ${\rm max}\langle\nu_{\rm eff}\rangle \gtrsim |k_\parallel|\vth{i0}$), with the collisionality decreasing with scale separation as ${\rm max}\langle\nu_{\rm eff}\rangle/\Omega_{i0} \propto (L_x/\rho_{i0})^{-1/6}$. (This is a purely empirical scaling for which we have no theory at present, although it is qualitatively consistent with larger scale separations resulting in a pressure anisotropy that overshoots the firehose threshold less before being regulated.) This translates to ${\max}\langle\nu_{\rm eff}\rangle/|k_\parallel|\vth{i0}\propto(L_x/\rho_{i0})^{5/6}$, a scaling that holds until ${\rm max}\langle\nu_{\rm eff}\rangle/|k_\parallel|\vth{i0} \sim \alpha\betaio$ at $L_x/\rho_{i0} \gtrsim 32000~(8000)$ for $\betaio=100~(25)$. This value is consistent with the asymptotic prediction \eqref{eqn:nueff} for the approximate effective collisionality required to regulate the IAW-driven pressure anisotropy through pitch-angle scattering alone (dotted line). Indeed, the $(\betaio,L_x/\rho_{i0})=(25,32000)$ simulation demonstrates that ${\rm max}\langle\nu_{\rm eff}\rangle/|k_\parallel|\vth{i0}$ saturates asymptotically at a value ${\approx}2\alpha\betaio$. The similar scaling of the $\betaio=25$ and $100$ runs, with the latter shifted to larger scale separations due to the larger value of $\alpha\betaio$, suggests that a run with $L_x/\rho_{i0} = 128000$ would also demonstrate saturation at ${\rm max}\langle\nu_{\rm eff}\rangle/|k_\parallel|\vth{i0} \approx 2\alpha\betaio$. Unfortunately, such a simulation is too computationally expensive to be practical at this time. Panel ($b$) demonstrates that, when time is normalized using $|k_\parallel|\vth{i0}$ and its origin is offset by the firehose-regulation time $t_{\rm f,reg}$, the effective collisionality in all $\betaio=100$ runs follows a similar evolution, with a sharp rise followed by a long exponential decay at a rate ${\approx}0.1|k_\parallel|\vth{i0}$. (The runs with $L_x/\rho_{i0}\ge 8000$ were too computationally expensive to be continued profitably for longer times.)

Finally, figure~\ref{fig:nuscan}($c$) presents the root-mean-square density fluctuation amplitude versus time for various combinations of $\betaio$ and $L_x/\rho_{i0}$. Larger scale separations cause a slight decrease in the rate at which the IAW is Landau damped and smaller values of $\alpha\betaio$ result in a larger initial damping rate, at least for $|k_\parallel|\vth{i0}t \lesssim 5$. Both trends are consistent with the scaling of ${\rm max}\langle\nu_{\rm eff}\rangle$ shown in panel ($a$): larger values of $\nu_{\rm eff}/|k_\parallel|\vth{i0}$ at larger scale separation reduce the potency of collisionless damping. After this time, the evolution of the IAW is roughly independent of both scale separation and $\alpha\betaio$.

In summary, figures \ref{fig:scan} and \ref{fig:nuscan} demonstrate that our findings are robust with respect to scale separation: IAWs with amplitudes satisfying \eqref{eqn:IAWlimit} produce enough pressure anisotropy to trigger firehose and mirror instabilities, which grow super-exponentially and then secularly, with the former generating an anomalous collisionality that is both large and lasting enough to make the plasma behave as though it were weakly collisional.

%
%
\section{Fluctuation-dissipation relation for self-sustaining IAWs}\label{sec:IAWLangevin}

While illustrative, the initial-value problem considered above is somewhat artificial. We thus consider the slightly more realistic problem of IAWs being stochastically and continuously driven and achieving a steady-state fluctuation-dissipation relation. Such a `plasma-kinetic Langevin' problem has been studied analytically in the linear regime (i.e., without feedback from pressure-anisotropy-driven instabilities) by \citet{kanekar15}. Those authors appended a stochastic momentum-injecting force, $2 v_\parallel a(t) \fMi/\vth{i0}^2$, to the right-hand side of the linearized ion-Vlasov equation \eqref{eqn:linvlasov}, where $a(t)$ had the two-time correlation function $\overline{a(t)a(t')} = \varepsilon(k) \vth{i0}^2 \delta(t-t')$ with $\varepsilon(k)$ being the energy-injection rate into wavenumber $k=k_\parallel$. They then solved for the mean square density-fluctuation amplitude in steady state, and found that it was given by
\begin{equation}\label{eqn:anjor}
    \overline{\left|\frac{\delta n(k)}{n_0}\right|^2} = \frac{2}{\upi} \frac{\varepsilon(k)}{|k|\vth{i0}} \biggl(\frac{T_{i0}}{T_e}\biggr)^2 \int^{+\infty}_{-\infty}\rmd\zeta \, \left|\frac{1+\zeta Z(\zeta)}{D(\zeta)}\right|^2 \doteq \frac{\varepsilon(k)}{2\gamma_{\rm eff}} \frac{T_{i0}}{T_e},
\end{equation}
where $D(\zeta)$ is given by \eqref{eqn:IAW} and the final equality implicitly defines the effective damping rate $\gamma_{\rm eff}$. The over-bar in \eqref{eqn:anjor} denotes a statistical-ensemble average with respect to the random forcing $a(t)$; we equate this average with a temporal average in the analysis below. In the cold-ion limit ($T_e/T_{i0}\rightarrow\infty$), $\gamma_{\rm eff}$ matches the Landau-damping rate given by \eqref{eqn:IAW_wk}; for $T_e/T_{i0} = 1$, $\gamma_{\rm eff} = 0.71 |k| \vth{i0}$ (see figure~4 of \citealt{kanekar15}).

%
%
\subsection{Pressure anisotropy and a nonlinear fluctuation-dissipation relation}

Fluctuation-dissipation relations of the form (\ref{eqn:anjor}) have the property that, if the effective damping rate decreases, the mean square density-fluctuation amplitude increases. Since the production of firehose and mirror instabilities by a non-linear IAW decreases the effective damping rate by interrupting Landau damping, the mean square density-fluctuation amplitude should increase accordingly whenever those instabilities are triggered. To determine what energy-injection rate would generate sufficiently large density fluctuations to destabilize the plasma to firehose and mirror, we calculate the mean square pressure anisotropy in steady state in a similar way as for \eqref{eqn:anjor}, using the $m_i v^2_\perp/2$ and $m_i v^2_\parallel$ moments of \eqref{eqn:linvlasov}; it is
\begin{equation}\label{eqn:langevinD}
    \overline{|\Delta(k)|^2} = \frac{2}{\upi} \frac{\varepsilon(k)}{|k|\vth{i0}} \biggl(\frac{T_{i0}}{T_e}\biggr)^2 \int^\infty_{-\infty} \rmd\zeta \, \biggl|\dfrac{\zeta Z(\zeta) (1-2\zeta^2) - 2\zeta^2}{D(\zeta)}\biggr|^2 .
\end{equation}
For $T_e/T_{i0}=1$, $\overline{|\Delta(k)|^2} = 3.33 \, \overline{|\delta n(k)/n_0|^2}$. In this case, using \eqref{eqn:anjor}, we estimate that having  $\varepsilon(k)/|k|\vth{i0} \ge 1.71/\beta^2_i$ would generate sufficiently large density fluctuations to cause $\overline{|\Delta(k)|^2} \ge (2/\beta_i)^2$.

We test this idea by forcing the parallel ion velocity stochastically and measuring the steady-state mean square density-fluctuation amplitude in a series of {\tt Pegasus} simulations, which are otherwise identical in their setup to those described in \S\ref{sec:IAWnumerics} ({\it viz.}, $\betaio=100$, $T_e=T_{i0}$, $L_x\times L_y=1000\rho_{i0}\times 50\rho_{i0}$). The forcing $\bb{a} = a(t,k)\ex$ is implemented as follows. At each simulation time step, the Fourier coefficients $a(t,k_x=\pm 2\upi/L_x)$ are generated from a Gaussian random field; their amplitudes are normalized by specifying a total energy-injection rate $\varepsilon$, whose magnitude is varied to result in root-mean-square density fluctuations than span the limit (\ref{eqn:IAWlimit}). The force is  time-decorrelated over $t_{\rm corr} = 20\upi\Omega^{-1}_{i0}\ll L_x/\rho_{i0}$ using an Ornstein--Uhlenbeck process.

%
%
\begin{figure}
    \centering
    \includegraphics[width=0.48\textwidth]{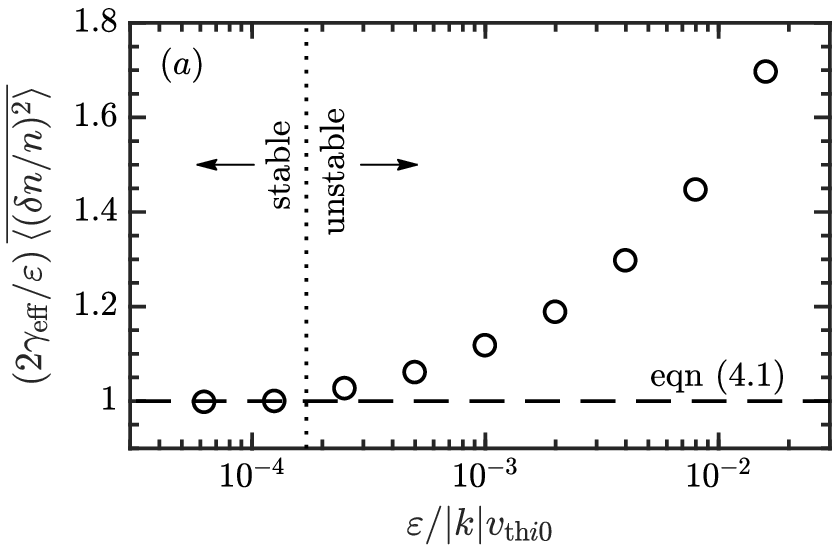}
    \includegraphics[width=0.48\textwidth]{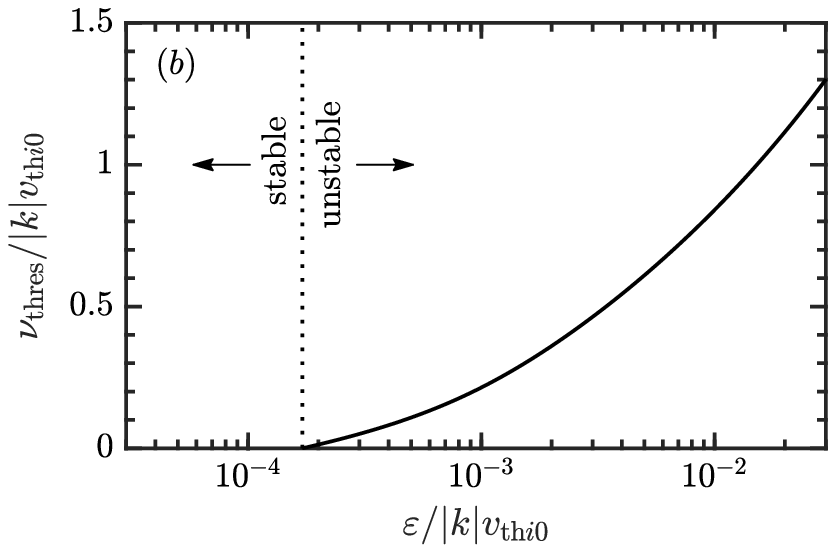}
    \caption{($a$) Mean square density-fluctuation amplitude for a given total energy-injection rate $\varepsilon$ in the IAW-Langevin problem with $\betaio=100$ and $L_x/\rho_{i0}=1000$ (circles), normalized to the linear prediction \eqref{eqn:anjor} without pressure-anisotropy-driven instabilities (dashed line). The increased fluctuation level that occurs when $\varepsilon$ is large enough to drive ${\rm max}|\Delta|>2/\betaio$ (dotted line; see \eqref{eqn:langevinD}) is due to the nonlinear suppression of collisionless damping by firehose and mirror instabilities. ($b$) Collision frequency $\nu_{\rm thres}$ needed to bound the steady-state pressure anisotropy in the linear Langevin problem (see \eqref{eqn:langevinD}) by the firehose instability threshold for $\betaio=100$.}
    \label{fig:langevin}
\end{figure}

Figure~\ref{fig:langevin}($a$) displays the mean square density-fluctuation amplitude measured in these simulations as a function of $\varepsilon$, normalized so that an amplitude of unity corresponds to \eqref{eqn:anjor}, the result without nonlinear feedback from firehose/mirror instabilities. Points to the left of the vertical dotted line are predicted using \eqref{eqn:langevinD} to be both firehose and mirror stable, i.e., such forcing results in a fluctuation amplitude whose ${\rm max}|\Delta|<2/\betaio$; those to the right are predicted to be unstable. As anticipated, the normalized fluctuation amplitude increases for $\varepsilon$ for which \eqref{eqn:IAWlimit} is satisfied, demonstrating that the effective damping rate $\gamma_{\rm eff}$ is nonlinearly reduced, {\em viz.}, $\gamma_{\rm eff} = \gamma_{\rm eff}(\varepsilon)$ -- a {\em nonlinear} fluctuation-dissipation relation.

%
%
\subsection{Effective collisionality and weakly collisional thermodynamics}

Another consequence of the reduction of $\gamma_{\rm eff}$ is that the plasma should exhibit behaviour more reminiscent of a weakly collisional (rather than collisionless) regime. Indeed, if we modify the dimensionless phase speed $\zeta \rightarrow \zeta + 3\imag\nu/|k|\vth{i0}$ in \eqref{eqn:langevinD} to account for pitch-angle scattering with collision frequency $\nu$, we may calculate from the linear Langevin problem the collision frequency needed for the steady-state pressure anisotropy to be bounded by the firehose instability threshold,  $\overline{|\Delta(k)|^2} \le (2/\betaio)^2$. This value, denoted $\nu_{\rm thres}$ and computed for a given $\varepsilon$, is shown in figure~\ref{fig:langevin}($b$); for $\varepsilon/|k|\vth{i0} \gtrsim 10^{-2}$, $\nu_{\rm thres} \sim |k|\vth{i0}$, a value large enough to make the otherwise collisionless plasma behave as a weakly collisional fluid.

%
%
\begin{figure}
    \centering
    \includegraphics[width=\textwidth]{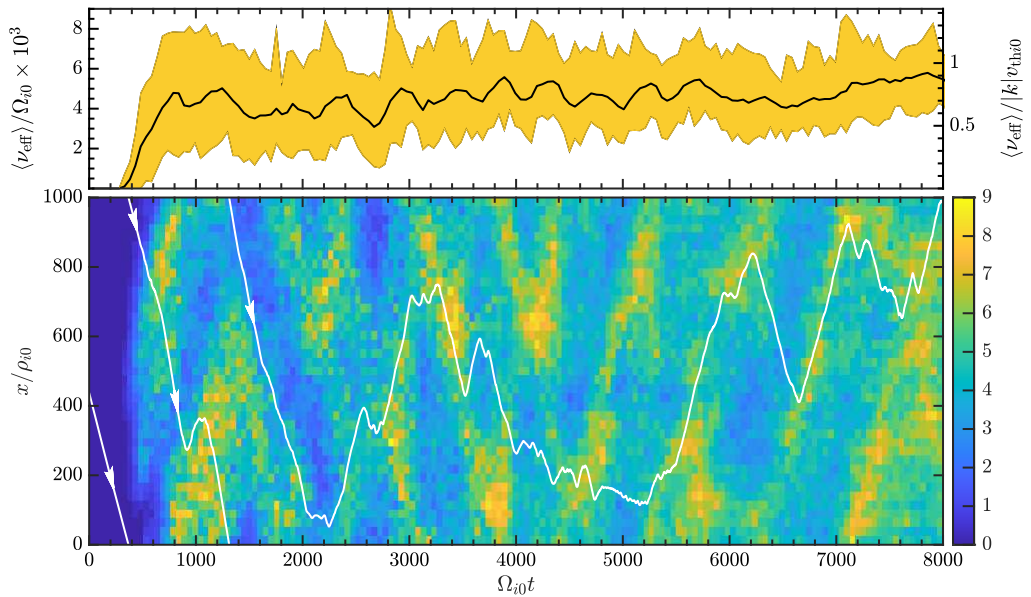}
    \caption{(top panel) Mean (black line) and spread (yellow region) of the effective collision frequency $\nu_{\rm eff}$, normalized using $10^{-3}\Omega_{i0}$ (left ordinate) and $|k|\vth{i0}$ (right ordinate), from the IAW-Langevin problem featured in figure~\ref{fig:langevin}($a$). (bottom panel) Space-time diagram of $\langle\nu_{\rm eff}\rangle_y/\Omega_{i0}\times 10^3$ (colour); the white line follows the trajectory of a representative particle.}
    \label{fig:langevin_coll}
\end{figure}

\begin{figure}
    \centering
    \includegraphics[width=\textwidth]{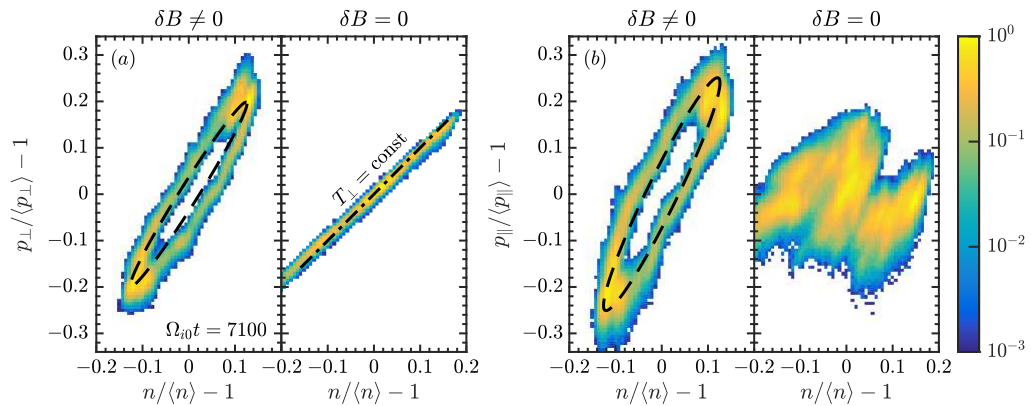}
    \caption{Distributions of fluctuating ($a$) perpendicular and ($b$) parallel pressures vs.~fluctuation density in two IAW-Langevin simulations: one allowing the generation of firehose/mirror instabilities ($\delta B\ne0$) and one with $\delta B=0$ set artificially. The dashed curves were obtained from \eqref{eqn:cgleqns} -- a weakly collisional fluid model based on the \citet{cgl56} equations -- using $\nu/|k|\vth{i0}=0.8$, the value of $\langle\nu_{\rm eff}\rangle/|k|\vth{i0}$ measured from the $\delta B\ne{0}$ simulation at $\Omega_{i0}t=7100$ (see figure~\ref{fig:langevin_coll}).}
    \label{fig:langevin_eos}
\end{figure}

That this plasma does, indeed, behave as a weakly collisional fluid is vividly illustrated in figures~\ref{fig:langevin_coll} and \ref{fig:langevin_eos}. The former shows the effective collision frequency $\nu_{\rm eff}$ computed following the same procedure as was used to obtain figure~\ref{fig:iaw_nueff}. The steady-state box-averaged collisionality $\langle\nu_{\rm eff}\rangle$ (black line) takes on values ${\approx}(4$--$6)\times 10^{-3}~\Omega_{i0} \approx (0.6$--$0.9)|k|\vth{i0}$, comparable to the maximum collisionality found in the corresponding initial-value problem at $L_x/\rho_{i0} = 1000$ (see figure~\ref{fig:iaw_nueff}).\footnote{A complementary Langevin simulation with $L_x/\rho_{i0}=2000$ returned $\langle\nu_{\rm eff}\rangle \approx (1.1$--$1.6)|k|\vth{i0}$, also comparable to the maximum collisionality found in the corresponding initial-value problem. The effective collision frequency shown in figure~\ref{fig:langevin_coll} is therefore an {\em under-estimate} of the asymptotic value of $\langle\nu_{\rm eff}\rangle/|k|\vth{i0}$, which we anticipate being ${\sim}(\delta n/n)\beta_i$ based upon theory (\S\ref{sec:IAWcoll}) and the measured scaling in the initial-value problem (figure~\ref{fig:nuscan}).} The yellow region indicates the spread in $\langle\nu_{\rm eff}\rangle$ about the mean, of width ${\approx}0.7|k|\vth{i0}$. The space-time evolution of $\nu_{\rm eff}$ (normalized to $10^{-3}\Omega_{i0}$) is shown in the bottom panel (colour), overlaid by the track (white line) of a representative particle that gets scattered and trapped intermittently by the kinetic instabilities. This data demonstrates that, in statistical steady state, the plasma has an anomalous collisionality sufficiently large to interrupt Landau damping.

This interruption makes a striking impact on the equation of state of the plasma. Figure~\ref{fig:langevin_eos} shows (in colour) joint distribution functions of the fluctuating perpendicular and parallel pressures and the fluctuating density measured within two IAW-Langevin simulations with $\varepsilon/|k|\vth{i0} = 1.6\times10^{-2}$: one with firehose and mirror instabilities allowed (labelled $\delta B\ne 0$, taken at $\Omega_{i0}t=7100$), and one purely electrostatic with Faraday's law of induction artificially turned off (labelled $\delta B = 0$). Figure~\ref{fig:langevin_eos}($a$) demonstrates that allowing kinetic instabilities ($\delta B\ne 0$) breaks adiabatic invariance in the plasma, {\it viz.}, it breaks $T_\perp={\rm const}$, a constraint shown to be strictly obeyed when $\delta B=0$ (the dot-dashed line). As a result, $\delta p_\perp$ is out of phase with $\delta n$. The dashed curve, which qualitatively captures the shape of this distribution, is a solution to the \citet{cgl56} fluid equations for a small-amplitude, monochromatic, parallel propagating ($k=k_\parallel$) sound wave in a weakly collisional plasma with collision frequency $\nu=0.8|k|\vth{i0}$ (as measured in the simulation at $\Omega_{i0}t=7100$) and isothermal electrons of temperature $T_e \doteq m_i c^2_s$:
\begin{subequations}\label{eqn:cgleqns}
\begin{align}
    \biggl( \pDD{t}{} + k^2_\parallel c^2_s \biggr) \frac{\delta n}{n} &= -\frac{k^2_\parallel \vth{i0}^2}{2} \frac{\delta p_{\parallel i}}{p_i} ,\\*
    \pD{t}{} \biggl( \frac{\delta p_{\perp i}}{p_i} - \frac{\delta n}{n} \biggr) &= -\nu \biggl( \frac{\delta p_{\perp i}}{p_i} - \frac{\delta p_{\parallel i}}{p_i} \biggr) ,\label{eqn:cgla}\\*
    \pD{t}{} \biggl( \frac{\delta p_{\parallel i}}{p_i} - 3 \frac{\delta n}{n} \biggr) &= -2\nu \biggl( \frac{\delta p_{\parallel i}}{p_i} - \frac{\delta p_{\perp i}}{p_i} \biggr) .\label{eqn:cglb}
\end{align}
\end{subequations}
In \eqref{eqn:cgla} and \eqref{eqn:cglb}, we neglected the conductive transport of heat along the (straight) field lines, but apparently without much consequence, as the dashed curve fits the measured $p_\perp$ distribution well enough to demonstrate the salient point: the equation of state of an otherwise collisionless plasma subject to self-excited velocity-space instabilities resembles that of a weakly collisional, non-conducting fluid. This weakly collisional solution also fits well the $p_\parallel$ distribution shown in figure~\ref{fig:langevin_eos}($b$) for $\delta B\ne 0$. In contrast, $p_\parallel$ in the collisionless plasma with $\delta B=0$ is roughly independent of density, a consequence of rapid free streaming of particles along the (unperturbed) magnetic-field lines.
    
%
%

%
%
\section{Discussion: Implications for astrophysical plasmas and theories of magneto-kinetic turbulence}

One of the key predictions of modern theories of Alfv\'{e}nic turbulence is that the compressive fluctuations are passively mixed by the Alfv\'{e}n-wave fluctuations \citep{lg01}, a feature which is manifest in the reduced MHD equations and which holds even in collisionless, magnetized plasmas \citep{schekochihin09,kunz15}. Moreover, the compressive fluctuations exert no influence on the Alfv\'{e}n-wave fluctuations, whose nonlinearities are solely due to the mutual advection of counterpropagating Alfv\'{e}n-wave packets. Given the findings of this paper, and those of its predecessors that focused on the interruption of nonlinear Alfv\'{e}n waves by their wave-generated pressure anisotropy \citep{squire16,squire17num,squire17}, it is clear that this tenet cannot hold at sufficiently large $\beta$. Pressure anisotropies driven by the Alfv\'{e}nic fluctuations with amplitudes $\delta B_\perp/B_0 \gtrsim \beta^{-1/2}$ can drive kinetic instabilities potentially capable of suppressing phase mixing and the consequent collisionless damping of ion-acoustic fluctuations. And ion-acoustic fluctuations with amplitudes $\delta n/n_0 \gtrsim \beta^{-1}$ (investigated in this paper) can develop pressure anisotropies large enough to interfere with the stability of Alfv\'{e}nic fluctuations. The mutual interactions between what are conventionally considered to be energetically decoupled cascades, and the impact of this coupling on the pillars of turbulence theory -- namely, the constant flux of energy and the locality of interactions in $k$ space -- now find themselves in our cross hairs.

In the meantime, let us speculate about the consequences of our results for the compressive fluctuations routinely measured in the turbulent solar wind. A particularly intriguing application is to the apparently fluid character of what ought to be Landau-damped kinetic fluctuations in that ostensibly collisionless environment. Using data from the {\em Wind} spacecraft in the solar wind at 1~au, \citet{verscharen17} found that a MHD description of ion-acoustic and non-propagating (i.e., pressure-balanced) modes fit much better with measurements of those fluctuations than did predictions based upon linear kinetic theory. This implies that some process is negating the effects of phase mixing. Our mechanism cannot of course explain this observational result for regions of the solar wind with $\beta\lesssim{1}$, which are necessarily firehose stable.\footnote{A possible explanation there is that stochastic echoes -- i.e., anti-phase-mixing modes -- arising from the nonlinear advection of the particle distribution by turbulent motions strongly suppress (on average) the phase mixing, a process recently suggested by \citet{schekochihin16} and \citet{meyrand19} to lead to `fluidization' of collisionless plasmas.} But, for those $\beta>{1}$ regions that have been measured to be constrained by the firehose and mirror instability thresholds and in which we see good evidence for the presence of firehose and mirror fluctuations \citep{kasper02,hellinger06,chen16}, we predict that wave-generated pressure anisotropy and the consequent production of firehose and mirror instabilities can interrupt collisionless damping and thus lead to compressive fluctuations that are more `fluid-like' than kinetic. A related suggestion had been made by \citet{verscharen16} in the context of the nearly collisionless solar wind: using linear theory, they argued that fluctuation-driven pressure anisotropy by long-wavelength compressive fluctuations and the consequent regulation of this anisotropy by pitch-angle-scattering kinetic instabilities could naturally explain the relative pressure isotropy of the $\beta>{1}$ solar wind. Our work supports this idea from first principles.

Another potential application is to wave propagation in the weakly collisional ICM. Sound waves generated by active galactic nuclei \citep[e.g.,][]{br19} and/or linear instabilities \citep[e.g.,][]{kempski20} may be important for the redistribution of energy throughout the ICM \citep[e.g.,][]{fabian03,ruszkowski04,fabian05,fabian17,zweibel18} and for observed density and temperature fluctuations in deep X-ray imaging of clusters \citep{zhuravleva19}. Our work implies that such waves, if they are large enough amplitude, propagate significantly more effectively than previously appreciated, enhancing the efficacy of sound wave energy redistribution throughout clusters. Although we have focused on the collisionless limit in this paper, the same physics applies in weakly collisional plasmas relevant to longer wavelength sound waves in clusters: waves with amplitudes $\delta n/n \gtrsim (\beta_i |k|\lambda_{\rm mfp,c})^{-1} $ will generate firehose and mirror instabilities, suppressing the damping of the wave. The implications of this for the thermodynamics of the ICM should be explored in future work.

In conclusion, the common wisdom that ion-acoustic fluctuations cannot propagate in a collisionless plasma with $T_i \sim T_e$ does not hold if $\delta n/n \gtrsim 2/\beta_i$. Given the high-$\beta$ conditions in many space and astrophysical plasmas, this result has broad relevance and opens several avenues for its application. An exploration of these avenues is now underway.

%
%
\section*{Acknowledgements}
Support for M.W.K.~was provided by the National Aeronautics and Space Administration (NASA) under Grant No.~NNX16AK09G issued through the Heliophysics Supporting Research Program. Additional support from DOE Award DE-SC0019047 and an Alfred P.~Sloan Research Fellowship in Physics is acknowledged. Support for J.S.~was provided by Rutherford Discovery Fellowship RDF-U001804 and Marsden Fund grant UOO1727, which are managed through the Royal Society Te Ap\=arangi. The work of A.A.S.~was supported in part by grants from UK STFC (ST/N0009/9/1) and EPSRC (EP/M022331/1 and EP/R034737/1). The work of E.Q.~was supported by DOE Award DE-SC0019046. High-performance computing resources were provided by: the Texas Advanced Computer Center at The University of Texas at Austin under grant number TG-AST130058; the NASA High-End Computing (HEC) Program through the NASA Advanced Supercomputing (NAS) Division at Ames Research Center; and the PICSciE-OIT TIGRESS High Performance Computing Center and Visualization Laboratory at Princeton University. This work used the Extreme Science and Engineering Discovery Environment (XSEDE), which is supported by NSF grant OCI-1053575. The completion of this work was facilitated by the generous hospitality and material support provided by the Wolfgang Pauli Institute in Vienna, as well as the expert re-write and optimization of the {\tt Pegasus} code (now {\tt Pegasus++}) by Lev Arzamasskiy, which made our scaling tests from $L_x/\rho_{i0}=8000$ up to $32000$ computationally feasible. The authors additionally thank Silvio Sergio Cerri for useful conversations and for kindly verifying some of our results obtained at $L_x/\rho_{i0}=500$ using the grid-based hybrid-Vlasov--Maxwell (HVM) code, and the referees for their constructive feedback.

\end{document}